\documentclass{pasa}

\usepackage{graphicx}	% Including figure files
\usepackage{amsmath}	% Advanced maths commands
\usepackage{amssymb}	% Extra maths symbols
\usepackage[separate-uncertainty=true]{siunitx}
\usepackage{physics}
\usepackage{booktabs}
\usepackage{nicefrac}
\usepackage{subcaption}

\DeclareSIUnit\beam{beam}
\DeclareSIUnit\jansky{Jy}
\DeclareSIUnit\parsec{pc}
\DeclareSIUnit\solarmass{M_\odot}
\DeclareSIUnit\deg{deg}
\DeclareSIUnit\pixel{pixel}
\DeclareSIUnit\gauss{G}
\DeclareSIUnit\counts{counts}
\DeclareSIUnit\arcmin{arcmin}

\newlength{\wdth}

\title[Searching for the Synchrotron Cosmic Web Again]{Searching for the Synchrotron Cosmic Web Again: \\
A replication attempt}

\author[Hodgson et al.]{
Torrance Hodgson$^{1,2}$\thanks{\href{mailto:torrance@pravic.xyz}{torrance@pravic.xyz}},
Melanie Johnston-Hollitt$^{2}$,
Benjamin McKinley$^{1, 3}$,
and Natasha Hurley-Walker$^{1}$
\\
\affil{$^{1}$International Centre for Radio Astronomy Research (ICRAR), Curtin University, 1 Turner Ave, Bentley, 6102, WA, Australia}
\affil{$^{2}$Curtin Institute for Computation, Curtin University, GPO Box U1987, Perth, 6845, WA, Australia}
\affil{$^{3}$ARC Centre of Excellence for All Sky Astrophysics in 3 Dimensions (ASTRO3D), Bentley, Australia}
}

\jid{PASA}
\doi{10.1017/pas.\the\year.xxx}
\jyear{\the\year}

\usepackage{aas_macros}
\usepackage{hyperref}

\hypersetup{colorlinks,citecolor=blue,linkcolor=blue,urlcolor=blue}

\begin{document}
\begin{frontmatter}
\maketitle

\begin{abstract}
We follow up on the surprising recent announcement by \citet{Vernstrom2021} of the detection of the synchrotron cosmic web. We attempt to reproduce their detection with new observations with the Phase II, extended configuration of the Murchison Widefield Array at \SI{118.5}{\mega \hertz}. We reproduce their detection methodology by stacking pairs of nearby luminous red galaxies (LRGs)---used as tracers for clusters and galaxy groups--contained in our low frequency radio observations. We show that our observations are significantly more sensitive than those used in Vernstrom et al\@., and that our angular sensitivity is sufficient. And yet, we make no statistically significant detection of excess radio emission along the bridge spanning the LRG pairs. This non-detection is true both for the original LRG pair catalogue as used in Vernstrom et al., as well as for other larger catalogues with modified selection criteria. Finally, we return to the original data sets used in Vernstrom et al\@., and find that whilst we clearly reproduce the excess X-ray emission from ROSAT, we are not able to reproduce any kind of broad and extended excess intercluster filamentary emission using the original \SI{118.5}{\mega \hertz} MWA survey data. In the interests of understanding this result, as part of this paper we release images of the 14 fields used in this study, the final stacked images, as well as key components of our stacking and modelling code.
\end{abstract}

\end{frontmatter}

%%%%%%%%%%%%%%%%%%%%%%%%%%%%%%%%%%%%%%%%%%%%%%%%%%

%%%%%%%%%%%%%%%%% BODY OF PAPER %%%%%%%%%%%%%%%%%%

\section{Introduction}

The `cosmic web' is a term used to evoke the structure of the Universe on the very largest of scales. In this model, dense clusters and galaxy groups are connected by diffuse filaments, forming a web like structure, and are interspersed with large, empty voids. Galaxy surveys have provided a strong empirical basis for this model \citep[e.g.][]{Baugh2004}, whilst cosmological simulations have shown it to be a consequence of gravitational instabilities acting upon small density perturbations in the early Universe \citep[e.g.][]{Cen1999,Dave2001}. These same simulations, however, have predicted something more: that up to 40\% of the baryonic content of the Universe resides along these filaments and around the periphery of clusters and galaxy groups, existing in a diffuse, highly ionised plasma, the so-called `warm-hot intergalactic medium' (WHIM). To date, the WHIM has proven difficult detect, however a number of recent works in this area have made increasingly convincing claims to have made detection \citep[see, e.g.:][]{Eckert2015,Nicastro2018,Tanimura2019,Tanimura2020,Macquart2020}.

This sparse, weakly magnetised WHIM is also predicted to have an associated radio signature, the `synchrotron cosmic web' \citep[see:][]{Brown2011,Vazza2015,Vazza2019}. As part of ongoing large-scale structure formation, cosmological simulations predict strong accretion shocks---having Mach numbers in the range $\mathcal{M} \sim 10 - 100$)---from in-falling matter along filaments and around the outskirts of clusters. These shocks should be capable of accelerating the electrons from within the WHIM to high energies by way of diffusive shock acceleration and this population of high energy electrons, in turn, are expected to radiate this energy as synchrotron emission as they interact with weak intercluster magnetic fields. In this way, the cosmic web is expected to have a synchrotron radio signature that traces out accretion shocks along its boundaries. The detection and confirmation of this radio emission would allow us to validate models of the large-scale structure of the Universe, as well as giving us insight into the poorly understood intercluster magnetic environments at the sites of these shocks.

This synchrotron cosmic web, however, is predicted to be extremely faint and has proven especially difficult to detect. Large-scale magnetohydrodynamic simulations by \citet{Vazza2019}, for example, point to a large population of radio-relic-like shocks well below the level of direct detection of any current or future radio telescopes. Only a small fraction of the very brightest knots in these shocks rise to the level of direct detection, and these are located principally around the most massive galaxy clusters \citep{Hodgson2021a}. \citet{Vacca2018} did in fact claim direct detection of numerous large-scale synchrotron sources associated with the cosmic web, but follow-up observations by \citet{Hodgson2020} rebuffed these claims. More recently, \citet{Govoni2019} claimed the detection of a radio `ridge' extending between clusters Abell 309 and 401, suggesting a diffuse, energetic and magnetised plasma extending between the merging clusters. Whilst this detection goes some way to validating our models, in this particular case the energy is provided by the merging dynamics of the clusters and is qualitatively different to the more general mechanisms of the synchrotron cosmic web. 

Other attempts to detect the synchrotron cosmic web have turned to statistical detection techniques to reveal faint emission sources buried beneath the noise of our current observations. Foremost among these methods is the cross-correlation technique. This method involves constructing a `best guess' kernel of the probable locations of cosmic web emission, and performing a radial cross-correlation of this kernel with the radio sky. A positive correlation at \SI{0}{\degree} offset is, in theory, indicative of a detection. In this way, this method hopes to reduce the noise by effectively integrating over a large enough area of the sky. Both \citet{Brown2017} and \citet{Vernstrom2017} used cross-correlation methods, and both were unable to make a definitive detection. In the former study, no positive correlation was detected. In the latter, \citet{Vernstrom2017} did indeed report a correlation, however the association of other sources such as active galactic nucleii (AGNs), star forming galxies (SFGs), and other cluster emission with their correlation kernel meant they were unable to attribute the peak at \SI{0}{\degree} to the cosmic web alone.

Recently, however, \citet{Vernstrom2021} (herein: V2021) have reported definitive detection of the synchrotron cosmic web using an alternative statistical method known as stacking. Their method attempted to measure the mean intercluster radio emission between pairs of close-proximity luminous red galaxies (LRGs). LRGs are known to have a strong association with the centre of clusters and galaxy groups \citep{Hoessel1980,Schneider1983,Postman1995}. Close-proximity pairs of LRGs therefore are likely to indicate close-proximity overdense regions of our Universe, and in turn we expect some fraction of these to be connected by a filament. Thus, V2021 stacked hundreds of thousands of low-frequency radio images of such pairs, which were rotated and rescaled so as to align all pairs to a common grid, before being averaged so as to find the mean image. After subtracting out a model for the LRG and cluster contribution, they reported finding excess emission with $> 5 \sigma$ significance along the length of the intercluster region. Moreover, this excess was detected by two independent instruments---in the Galactic and Extragalactic All-sky MWA\footnote{Murchison Widefield Array \citep[MWA;][]{Tingay2013}} survey \citep{Wayth2015,HurleyWalker2017} and by the Owens Valley Radio Observatory Long Wavelength Array \citep[OVRO-LWA;][]{Eastwood2018}---and across four frequencies ranging from \SIrange[range-phrase=--,range-units=single]{73}{154}{\mega \hertz}. A null test, formed by stacking physically distant LRG pairs for which we do not expect a connecting filament to exist, returned no excess emission. After excluding multiple alternative explanations, V2021 suggested the most likely explanation for this excess intercluster signal was the cosmic web itself. 

The result reported in V2021 is convincing, but it is also surprising. Previous intercluster magnetic field estimates provided upper limits on the order of just a few nG \citep[e.g.][]{Pshirkov2016,OSullivan2019,Vernstrom2019}. However, the reported excess emission supports intercluster magnetic field strengths averaging \SIrange[range-phrase=--,range-units=single]{30}{60}{\nano \gauss}, and moreover these estimates are strictly a lower limit as some significant fraction of stacked pairs will not in fact be connected by a filament. More recent follow-up work by \citet{Hodgson2021b}, which stacked a simulated radio sky---including cosmic web emission---from the FIlaments and Galactic RadiO \citep[FIGARO;][]{Hodgson2021a} simulation, failed to reproduce excess intercluster emission. With perfect knowledge of their simulated sky, this work stacked the known locations of dark matter halos rather than LRGs. They reported excess emission being detected on the immediate interior of halo pairs, associated with asymmetric accretion shocks onto clusters and galaxy groups, but no detectable emission along the true intercluster region. This work also explored the role of other contaminating sources, such as AGN, SFG, and radio halo populations, as well as the effect of sidelobes from the dirty interferometric beam, finding none of these to be significant. The discrepancy between V2021 and these simulated results, which build on our current best simulations of the cosmic web, remain difficult to explain.

Given the importance of the result of V2021, in this present study we attempt to reproduce and corroborate their result. We do so using the upgraded MWA Phase II instrument \citep{Wayth2018}, observing at \SI{118}{\mega \hertz}, and take advantage of improvements to calibration and imaging pipelines that have appeared since the original GLEAM survey. We image 14 fields spanning the same LRG pairs as used in V2021, which we then stack using independent stacking and modelling pipelines. Our aim is to closely adhere to the methodology used in V2021 whilst seeking to measure the excess intercluster emission more accurately, thanks to the improved noise characteristics of our observations.

Throughout this paper, we assume a $\Lambda$CDM cosmological model, with density parameters $\Omega_{\text{BM}} = 0.0478$ (baryonic matter), $\Omega_{\text{DM}} = 0.2602$ (dark matter), and $\Omega_{\Lambda} = 0.692$, and the Hubble constant $H_0 =$ \SI{67.8}{\kilo \metre \per \second \per \mega \parsec}. All stated errors indicate one standard deviation.

\section{Luminous red galaxy pairs}

Only a few thousand clusters are currently catalogued with robust X-ray or Sunyaev–Zeldovich measurements \citep[e.g.][]{MCXC2011,PlanckCollab2016}. This number is much smaller than the expected number of clusters and galaxy groups \citep[e.g.][]{Wen2018}, and is also too few to be useful for our present purposes as we do not expect the faint, intercluster emission to become detectable above our field noise after stacking so few images. Instead, as with V2021, we turn to using LRGs as a proxy for such overdense regions. LRGs are massive, especially luminous early-type galaxies, and are closely associated with overdense regions of the Universe. This association, however, comes with some caveats as explored in detail in \citet{Hoshino2015}. To summarise briefly here, in the first instance not all massive clusters have a LRG as their central galaxy: for the most massive clusters the probability of this association peaks at 95\%, however this association steeply drops off for lower mass systems, reaching just 70\% for clusters of mass $M_{200} =$ \SI{E14}{\solarmass}. An additional error introduced from the brightest LRGs located in clusters that do not align with the cluster centre, and this `miscentred' fraction is substantial at \SIrange[range-phrase=--,range-units=single]{20}{30}{\percent}. Nonetheless, as with V2021, these caveats are acceptable given the vastly greater number of potential clusters that LRGs allow us to identify. It does, however, mean that any excess emission attributed to intercluster regions are strictly lower limits.

As with V2021, we use the LRG catalogue from \citet{Lopes2007}. This catalogue incorporates approximately $1.4$ million LRGs extracted from the fifth data release of the Sloan Digital Sky Survey \citep{York2000}, out to a redshift of $z < 0.70$. \citet{Lopes2007} have used an  empirical based method to calculate spectroscopic redshifts for this population from just three bands (\textit{gri}), with an estimated error $\sigma = 0.027$ for $z < 0.55$, and $\sigma = 0.040$ up to $z = 0.70$.

\begin{figure}
    \centering
    \includegraphics[width=\linewidth,clip,trim={0 0cm 0 0cm}]{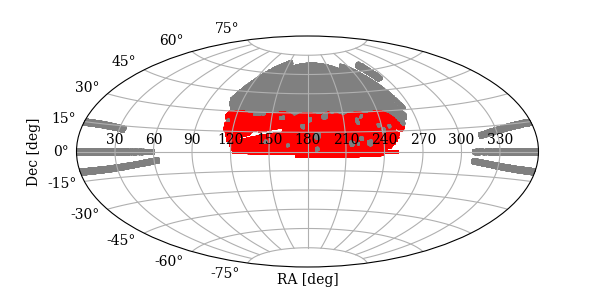}
    \caption{LRG pair distribution on the sky. The red points indicate pairs used in our stacks, whilst the grey are those pairs either outside our field or are within an exclusion zone.}
    \label{fig:lrgmap}
\end{figure}

In V2021, a list of LRG pairs was calculated that met the following set of conditions. First, the separation between the pairs was less than \SI{15}{\mega \parsec}. The metric used was the comoving distance, and this condition also included a lower bound of \SI{1}{\mega \parsec} (T. Vernstrom, personal communication). Additionally, the pairs were required to have an angular separation on the celestial sphere in the range \SI{20}{\arcminute}~$< \theta <$~\SI{180}{\arcminute}. We find 1,078,730 such valid pairs, of which 601,435 ultimately overlap with our fields, and label this catalogue `Max \SI{15}{\mega \parsec}'. In \autoref{fig:lrgmap} we show the location of the LRG pair population on the celestial sphere, with those in red overlapping with our fields.

V2021 reported finding just 390,808 pairs that satisfied these conditions. As it turns out, this reduced catalogue was the result of a bug in their code (T. Vernstrom, personal communication). Therefore, to compare like for like, we additionally include this abridged catalogue as `LRG-V2021'.

We also provide two additional catalogues of LRG pairs with differing selection criteria. In the first, we reduce the maximum spatial separation to \SI{10}{\mega \parsec} (`Max \SI{10}{\mega \parsec}'), of which there are 270,458 (153,433 overlapping) entries. The motivation for this catalogue arises from our expectation that intercluster emission should be brighter for cluster pairs in closer proximity to each other where they are more likely to be interacting, possibly triggering pre- or post-merger shocks known to produce synchrotron emission. We also include a final catalogue with modified angular constraints, such that the minimum and maximum angular separations are shifted to \SI{15}{\arcminute} and \SI{60}{\arcminute}, respectively (`Max \SI{60}{\arcminute}'). This catalogue contains 824,773 (436,899 overlapping) entries, and is motivated by concerns about resolving out large-scale angular structures, which is an aspect we discuss later in \autoref{sec:resolvingout}.

% In addition to a catalogue of LRG pairs using this criteria, we provide two more with the sole modification that the comoving separation is reduced to \SI{10}{\mega \parsec} (`Max \SI{10}{\mega \parsec}') and \SI{7.5}{\mega \parsec} (`Max \SI{10}{\mega \parsec}'), each containing 270,458 (244,647 overlapping) and 109,495 (99,208) pairs, respectively. These additional catalogues are motivated by our expectation that intercluster emission should be brighter for cluster pairs in closer proximity to each other where they are more likely to be interacting, possibly triggering pre- or post-merger shocks known to produce synchrotron emission.

\begin{figure*}
    \centering
    \includegraphics[width=\linewidth]{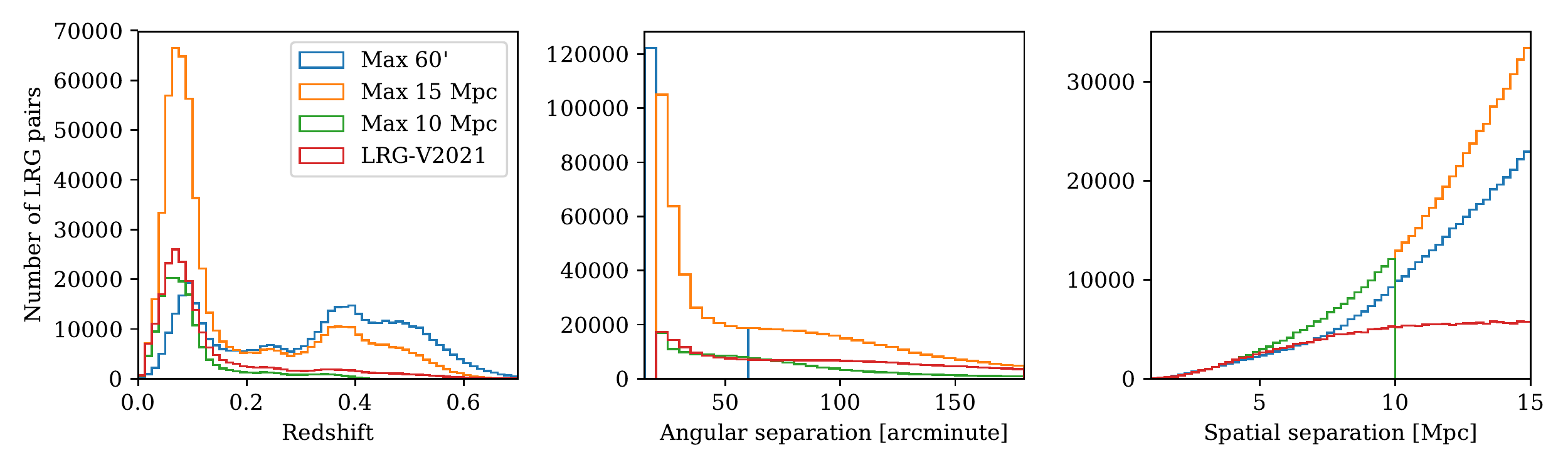}
    \caption{The LRG pair distributions by redshift (left), angular separation (centre), and spatial separation (right), for each of the LRG pair catalogues used in our stacks.}
    \label{fig:lrgpairs}
\end{figure*}

% \begin{table}
%     \centering
%     \begin{tabular}{ccccc} \toprule
%         & \textbf{Max 15 Mpc} & \textbf{Max 10 Mpc} & \textbf{Max 60$'$} & \textbf{LRG-V2021} \\ \midrule
%         $\theta$ & \SI{20}{\arcminute}~$< \theta <$~\SI{180}{\arcminute} & \SI{20}{\arcminute}~$< \theta <$~\SI{180}{\arcminute} &  \SI{20}{\arcminute}~$< \theta <$~\SI{180}{\arcminute} &  \SI{20}{\arcminute}~$< \theta <$~\SI{180}{\arcminute} \\
%         N & 969,313 & 244,647 & 726,186 & 354,154 \\
%         $\expval{z}$ & 0.185 & 0.099 & 0.33 & 0.14 \\
%         $\expval{\Delta \theta}$ & \SI{67}{\arcminute} & \SI{66}{\arcminute} & \SI{28}{\arcminute} & \SI{83}{\arcminute} \\
%         $\expval{\Delta R}$ & \SI{11.6}{\mega \parsec} & \SI{7.6}{\mega \parsec} & \SI{11.5}{\mega \parsec} & \SI{10.3}{\mega \parsec} \\ \bottomrule
%     \end{tabular}
%     \caption{LRG pair statistics comparison between each of the LRG pair catalogues. We show the number of overlapping LRG pairs, their mean redshift, their mean angular separation, and their mean spatial separation, respectively.}
%     \label{tab:halostats}
% \end{table}

\begin{table*}
    \centering
    \begin{tabular}{rccccccc} \toprule
        & \textbf{Spatial criteria} & \textbf{Angular criteria} & \textbf{N} & $\mathbf{\expval{z}}$ & $\mathbf{\expval{\Delta r}}$ & $\mathbf{\expval{\Delta \theta}}$\\
        & [Mpc] & [arcminute] & & & [arcminute] & [Mpc] \\ \midrule
        \textbf{Max 15 Mpc} & $1 < r < 15$ & $20 < \theta < 180$ & 601,435 (1,078,730) & 0.18 & 11.5 & 67.8 \\
        \textbf{Max 10 Mpc} & $1 < r < 10$ & $20 < \theta < 180$ & 153,433 (270,458)& 0.098 & 7.6 & 65.8 \\
        \textbf{Max 60$'$} & $1 < r < 15$ & $15 < \theta < 60$ & 436,899 (824,773) & 0.32 & 11.4 & 28.5 \\
        \textbf{LRG-V2021} & $1 < r < 15$* & $20 < \theta < 180$* & 219,684 (390,808) & 0.14 & 10.3 & 83.8 \\ \bottomrule
    \end{tabular}
    \caption{LRG pair statistics comparison between each of the LRG pair catalogues. We show the spatial and angular selection criteria for each catalogue, the number of LRG pairs that overlap with our fields (and the total pairs), their mean redshift, their mean angular separation, and their mean spatial separation, respectively. Spatial distances use a comoving metric. *Due to an error in the work of V2021, the LRG-V2021 catalogue is an incomplete catalogue that nonetheless adheres to these ranges.}
    \label{tab:lrgpairs}
\end{table*}

In \autoref{fig:lrgpairs}, we show the redshift, angular separation and spatial separation distributions of each of these LRG catalogues, for those LRG pairs that overlap with our fields and are included in our stacks. Note the double peak structure present in the redshift distribution of the Max 15 Mpc and Max \SI{60}{\arcminute} catalogues: this is a function of the underlying distribution of the LRG catalogue which exhibits a small peak around $z = 0.08$ and much larger peak around $z = 0.5$, combined with the effect at increasing redshifts of the duel constraints of the minimum angular separation and the maximum spatial separation. In the case of the \SI{15}{\mega \parsec} criterion, we find a mean redshift of $\expval{z} = 0.185$, a mean separation of $\expval{r} =$~\SI{11.6}{\mega \parsec}, and a mean angular separation of $\expval{\theta} =$~\SI{67}{\arcminute}. The mean values for the other LRG pair catalogues are provided in \autoref{tab:lrgpairs}.

\section{Observations \& data processing}

\subsection{Data selection}

The original GLEAM survey, which was used in V2021, was observed using the MWA Phase I \citep{Tingay2013}. This consisted of 128 tiles positioned to give a maximum baseline of approximately \SI{3}{\kilo \metre} when observing at zenith, and a large number of baselines under \SI{100}{\metre}; when observing near the horizon these baselines are significantly foreshortened. The upgrade to the MWA Phase II \citep{Wayth2015} in late 2017 was primarily a reconfiguration of the tile positions: the same 128 tiles were positioned to give an increased maximum baseline of almost \SI{6}{\kilo \metre} as well as a much smoother distribution of baselines. The effect of these changes was to give Phase II almost twice the resolution as well as a better behaved dirty beam with reduced sidelobes, whilst otherwise leaving the point source sensitivity unchanged. Sidelobe confusion is a major source of noise in Phase I observations, whereas in Phase II observations the higher resolution allows much deeper cleaning, which has flow on effects to further reduce image noise even when accounting for the resolution difference. In the observations used in this study, we take advantage of these improved characteristics of the MWA Phase II.

We have drawn our observations from those made in preparation for the upcoming GLEAM-X survey \citep{HurleyWalker2017b}. GLEAM-X has observed the sky at frequencies ranging from \SIrange[range-phrase=--,range-units=single]{72}{231}{\mega \hertz} in short duration `snapshots' of approximately 2 minutes. These observation runs are typically observed at a fixed pointing in a `drift scan' mode, where the celestial sphere is allowed to freely rotate through the primary beam.

We have identified 14 fields to image that best span the LRG population. These fields are centred at declinations of $\delta = \{+2^\circ, +18^\circ\}$ and spanning the right ascension range of $120^\circ \leq \alpha \leq 240^\circ$ at intervals of \SI{20}{\degree}. From the archive of GLEAM-X observations, we filter for snapshots observed at \SI{118.5}{\mega \hertz} and where their pointing centres are located near the centre of these 14 fields, with a tolerance $\alpha \pm 5^\circ$ and $\delta \pm 3^\circ$. There are 512 observations that match this criteria, made during runs in February--March 2018, May--June 2018, January--February 2019, and March 2019. After calibration, imaging, and quality control checks, however, this number is reduced to 291 snapshots, constituting approximately 10 hours of observations. In \autoref{tab:fields} we tabulate these 14 fields and some of their properties.

\begin{table*}
    \centering
    \begin{tabular}{ccccccp{0.35\linewidth}} \toprule
        \textbf{ID} & \textbf{RA} & \textbf{Dec} & \textbf{Snapshot}s & \textbf{Noise} & \textbf{Model deviation} & \textbf{Notes} \\
        & [deg] & [deg] & & [\SI{}{\milli \jansky \per \beam}] & $\mu$ | $\sigma$ [dex] & \\ \midrule
        1 & 120 & 3 & 19 & 8.2 & 0.000 / 0.025 & \\
        2 & 140 & 3 & 35 & 5.8 & 0.001 / 0.021 & Hydra A present in field, peaking at $\sim$\SI{260}{\jansky \per \beam}. \\
        3 & 160 & 3 & 28 & 7.1 & 0.000 / 0.022 & Affected by sidelobes from Virgo A\\
        4 & 180 & 3 & 35 & 6.6 & -0.002 / 0.028 & Virgo A present in field peaking at $\sim$\SI{526}{\jansky \per \beam}, and second bright source present 3C 273 peaking at $\sim$\SI{105}{\jansky \per \beam}.\\
        5 & 200 & 3 & 31 & 8.8 & 0.000 / 0.022 & Large-scale sidelobe pattern present from Centaurus A which is positioned south out-of-field in sidelobe of the primary beam. Virgo A also present in field.\\
        6 & 220 & 3 & 15 & 8.4 & 0.001 / 0.026 & \\
        7 & 240 & 3 & 11 & 10.0 & -0.001 / 0.022 & Hercules A present in field peaking at $\sim$\SI{377}{\jansky \per \beam}\\
        8 & 120 & 18 & 11 & 8.0 & -0.002 / 0.027 & \\
        9 & 140 & 18 & 17 & 6.2 & 0.000 / 0.026 & Large-scale sidelobe pattern from Virgo A in south, out-of-field. \\
        10 & 160 & 18 & 20 & 6.4 & -0.001 / 0.030 & \\
        11 & 180 & 18 & 24 & 6.8 & -0.001 / 0.035 & Both Virgo A and 3C 273 present in field.\\
        12 & 200 & 18 & 19 & 9.5 & -0.004 / 0.037 & Virgo A present, as well as large-scale sidelobe pattern from Centaurus A, positioned south out-of-field.\\
        13 & 220 & 18 & 11 & 11.1 & -0.003 / 0.036 & \\
        14 & 240 & 18 & 7 & 14.6 & 0.001 / 0.031 & \\
        % 15* & 210 & 18 & 10 & 13.6 & 1.00 / 0.12 & Weak, large-scale sidelobes from out-of-field sources Virgo A and Centaurus A.\\ 
        \bottomrule
    \end{tabular}
    \caption{A summary of the 14 fields imaged, observed by the MWA Phase II instrument at \SI{118}{\mega \hertz}. The fields span the right ascension range \SI{120}{\degree} to \SI{240}{\degree} in \SI{20}{\degree} increments, at declinations of \SI{3}{\degree} and \SI{18}{\degree}. We indicate the number of \SI{112}{\second} duration snapshots used in each field mosaic, and the resulting noise at the centre of the field. The model deviation describes the ratio of the measured flux density of sources after performing source finding, in comparison to the original calibration sky model; the $\mu$ term describes the mean values of these ratios, whilst $\sigma$ shows the standard deviation of these ratios.}
    \label{tab:fields}
\end{table*}

\subsection{Data processing}

All observations are centred at \SI{118.5}{\mega \hertz}, of \SI{112}{\second} duration, spanning a bandwidth of \SI{30.72}{\mega \hertz}, and correlated at a resolution of \SI{10}{\kilo \hertz} and \SI{0.5}{\second}. This is further averaged to \SI{40}{\kilo \hertz} and \SI{4}{\second} prior to calibration and imaging to ease data storage and processing requirements. All subsequent data processing occurs on a per-snapshot basis until final mosaicing.

Calibration is performed using an in-field radio sky model. This sky model has been constructed in preparation for the GLEAM-X survey, and is principally based on the GLEAM sky catalogue. It does, however, include a number of additional sources, including better models for the so-called `A-team' of extremely bright radio sources, such as Hydra A, Virgo A, Hercules A, and Centaurus A, all of which populate our fields. The GLEAM sky catalogue is known to have an error of \SI{8.0(5)}{\percent} up to declination \SI{18.5}{\degree}, and an uncertainty of \SI{11(2)}{\percent} for more Northern declinations. We calibrate on all sources from this sky model that are within a \SI{20}{\degree} radius of our field centre, and having a primary beam-attenuated apparent flux density of at least \SI{700}{\milli \jansky}. These sources are predicted into the visibilities using the full embedded element primary beam model \citep{Sokolowski2017}. 

Calibration is performed using the updated MWA calibrate tool \footnote{See \url{https://github.com/torrance/MWAjl/}.}, which finds a full Jones matrix solution for each antenna, independently for each pair of channels, and with the solution interval set to the duration of the snapshot. Baselines longer than approximately \SI{3.7}{\kilo \metre} are excluded from consideration during calibration, since these baselines are increasingly sensitive to angular scales of a higher resolution than the original GLEAM catalogue. Calibration solutions are visually inspected, and any antennae which have failed to well converge are flagged at this time. No self-calibration is performed, as in practice we have found this to be unnecessary. 

Imaging is performed using \texttt{wsclean} \citep{Offringa2014}. We weight baselines using the Briggs formulation, with a robustness factor of +1; additionally baselines smaller than $15 \lambda$, which are sensitive to emission on angular scales larger than \SI{3.8}{\degree}, are excluded to avoid any kind of large-scale contamination from Galactic emission. Cleaning is then performed down to a threshold that depends on two factors: cleaning continues until first the `auto-mask' threshold is reached, which is set at a factor of 3 times the residual map noise, and then cleaning continues only on those pixels previously cleaned down to the `auto-threshold' limit, which we set as the estimated residual map noise. Typical values for the residual map noise of these individual snapshots is around \SIrange[range-phrase=--,range-units=single]{15}{20}{\milli \jansky}. During imaging, we split the \SI{30.72}{\mega \hertz} band into four equally sized channels to account for the typical flux density changes of sources over this frequency range due both to intrinsic properties and beam attenuation. We do, however, perform joint-channel cleaning, where clean peaks are chosen based on a full-bandwidth mean map, and the peak value is estimated using a linear fit across each output channel. Note that whilst \texttt{wsclean} does have multiscale clean functionality, we have chosen not to use this, so that any faint, extended emission sources \textit{remain} in the residual maps after cleaning. We image and clean instrumental polarizations (e.g.\@ XX, XY, YX, YY) independently, which is important since sources at this low elevation become strongly polarised as a result of the primary beam. These instrumental polarization images are later combined based on the primary beam model to produce Stokes I images. Finally, after imaging is completed, we keep both the restored and residual Stokes I images for each snapshot for later processing: the restored map is used to verify and correct field calibration, whilst the residual map provides us with a point-source subtracted map to be ultimately used in stacking, without the need for complex wavelet subtraction techniques as used in V2021.

As a first order effect of ionospheric electron density variations, we observe direction-dependent shifts in the apparent position of radio sources, and these effects become increasingly strong at the low frequencies observed by the MWA. Without resolving this positional error, we not only risk introducing astrometric errors, but additionally sources in the final mosaic can appear blurred and point sources have a peak to integrated flux density ratio that is less than unity. To resolve this, typical MWA workflows make image based corrections to `warp' the image, and align the apparent position of sources with their position in the sky model (for example, see \citealp{HurleyWalker2018}). We follow this method by first source finding on the restored image using \texttt{Aegean} \citep{Hancock2018}, and cross-matching these sources with our sky model. We include only those sources that are isolated by at least \SI{1}{\arcminute} radius from any other sky model source to avoid any ambiguous matches, and as a quality control check we require at least 200 cross-matches in a snapshot or else it is discarded. Then by measuring the angular offset of apparent position to that of the sky model, we interpolate across these deviations and thus warp the image to correct for this effect. 

In an effort to match the sensitivity to extended emission of the MWA Phase II instrument to that of its Phase I counterpart, we proceed by convolving both the restored and residual images. At \SI{118.5}{\mega \hertz}, a typical dirty beam size at Briggs +1 weighting has major and minor axes of approximately \SI{2.3 x 1.8}{\arcminute}, whilst this size varies significantly by declination due to the foreshortening effect of the array at low elevations. We use \texttt{miriad} \citep{Sault1995} to convolve each snapshot to a circularised resolution of \SI{3}{\arcminute}, defined at zenith. We discuss the effects of this convolution step, and our sensitivity to extended emission, in \autoref{sec:resolvingout}.

% We use the `shiftback' functionality of \texttt{wsclean} to image individual snapshots on a slant orthographic projection (`SIN') with the projection origin approximately at zenith; this has the benefit of reducing the computational cost of imaging by decreasing the number of $w$ layers \texttt{wsclean} must use, but additionally it allows for cleaning to converge easily since the dirty beam skews with altitude in the same way as the projection.

Our snapshots are ready to be stacked and mosaiced. To do this we must first ensure all images are on the same projection, which are presently in a slant orthographic projection (`SIN') with the projection origin at each snapshot's zenith. To minimise reprojection errors, which can be significant, we choose to reproject each snapshot onto the mean projection shared amongst the snapshots for a particular field, leaving the SIN projection origin approximately at the MWA zenith. We additionally mask the region within \SI{15}{\degree} of the horizon for each snapshot, as these low-elevation observations are subject to significant errors. With these steps completed, we perform the weighted mean of all snapshots, with the weight based on the estimated local map noise $\sigma$, as $\nicefrac{1}{\sigma^2}$.\footnote{The local noise map is calculated using the median absolute deviation from the median (MADM) applied to a residual image that has \textit{not} been primary beam corrected. We choose to use this image for our noise estimation as it has had bright sources removed and, prior to beam correction, the noise does not vary spatially, thus allowing for the easy calculation of a global value. We then apply a beam correction to this constant noise map so as to obtain an estimate of the local noise map, which varies spatially as a function of the primary beam.} A final quality check is included during this mosaicing step, whereby any snapshot with a map noise in excess of \SI{35}{\milli \jansky \per \beam} (increased to \SI{45}{\milli \jansky \per \beam} for field 14) is discarded. In this way, we create mosaics of the residuals, the restored images, as well as the estimated noise.

\begin{figure}
    \centering
    \vspace{-1.2cm}
    \includegraphics[width=\linewidth,clip,trim={0.4cm 0.4cm 0.2cm 0.2cm}]{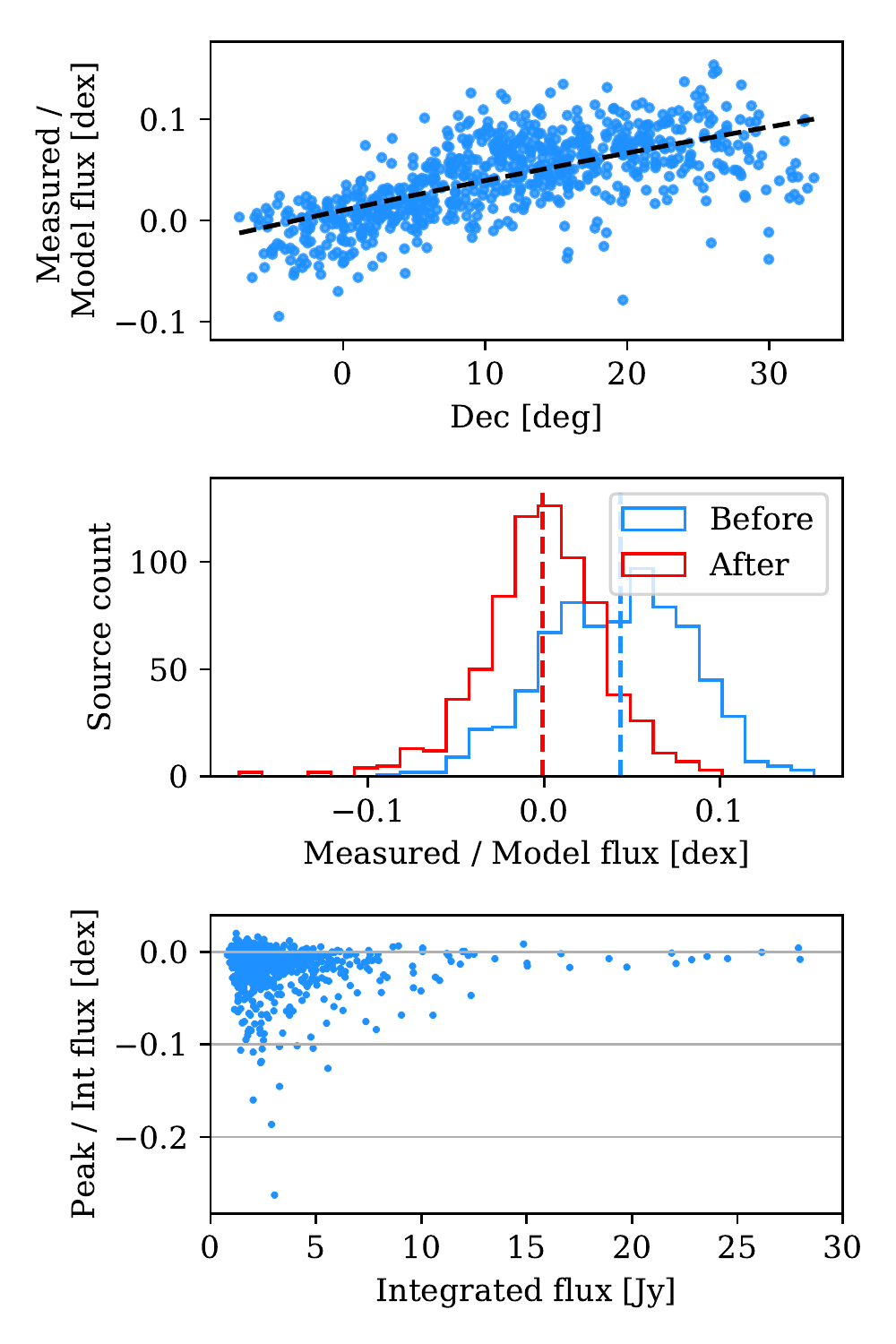}
    \caption{Example calibration diagnostics for field 10, showing the declination correction. The 731 measured sources are compared to the calibration model. Prior to correction, their ratio has mean 0.044 dex and standard deviation 0.044 dex; after correction the mean becomes -0.001 dex and standard deviation 0.030 dex.  \textit{Top:} The measured to model flux density ratio, as a function of declination. The dashed line indicates the fit which is later used as an image-based correction. \textit{Centre:} The distribution of measured to model flux density ratios for all 731 sources prior to (blue) and after (red) correction. Note in this case, the simple declination correction resolves the initial bimodal distribution. \textit{Bottom:} The measured ratio of peak to integrated flux density, showing peak and integrated flux density of point sources are very nearly identical.}
    \label{fig:deccorrection}
\end{figure}

Finally, we verify our calibration by source finding on the final mosaic and comparing the measured flux density to the sky model flux density. As reported by \citet{HurleyWalker2017}, we observe a declination-dependent flux density error. In \autoref{fig:deccorrection}, we present the kind of diagnostic used to check the flux density values for each field. For example, the top panel of this figure shows this error across field 10, showing the measured to model integrated flux density ratio increases from approximately unity to as high as 1.3 times at declination \SI{+30}{\degree}. To correct for this effect, we model this error as a simple linear function of declination, as depicted by the dashed black line, and scale the image accordingly. The centre panel in \autoref{fig:deccorrection} shows the effect of this correction: the mean flux density density ratio is reduced from 0.044 dex to -0.001 dex; the spread of ratios is reduced from a standard deviation of 0.0436 dex to 0.030 dex; and in this particular instance, the apparently bimodal distribution is corrected to appear much more normally distributed. As shown in \autoref{tab:fields}, the mean flux density ratio across all fields is very nearly 0 {dex} after this correction, whilst the standard deviation lies in the range \SIrange[range-phrase=--,range-units=single]{0.021}{0.037}{dex}. These values are well within the stated errors of the GLEAM sky model. As a final sanity check, we also show in the lower panel of \autoref{fig:deccorrection} the ratio of peak to integrated flux density, where we can observe a good clustering of values around unity, suggesting that our ionospheric corrections are satisfactory.

\begin{figure*}
    \centering
    \vspace{-1.6cm}
    \includegraphics[width=\linewidth,clip,trim={1.8cm 1.5cm 2.5cm 0cm}]{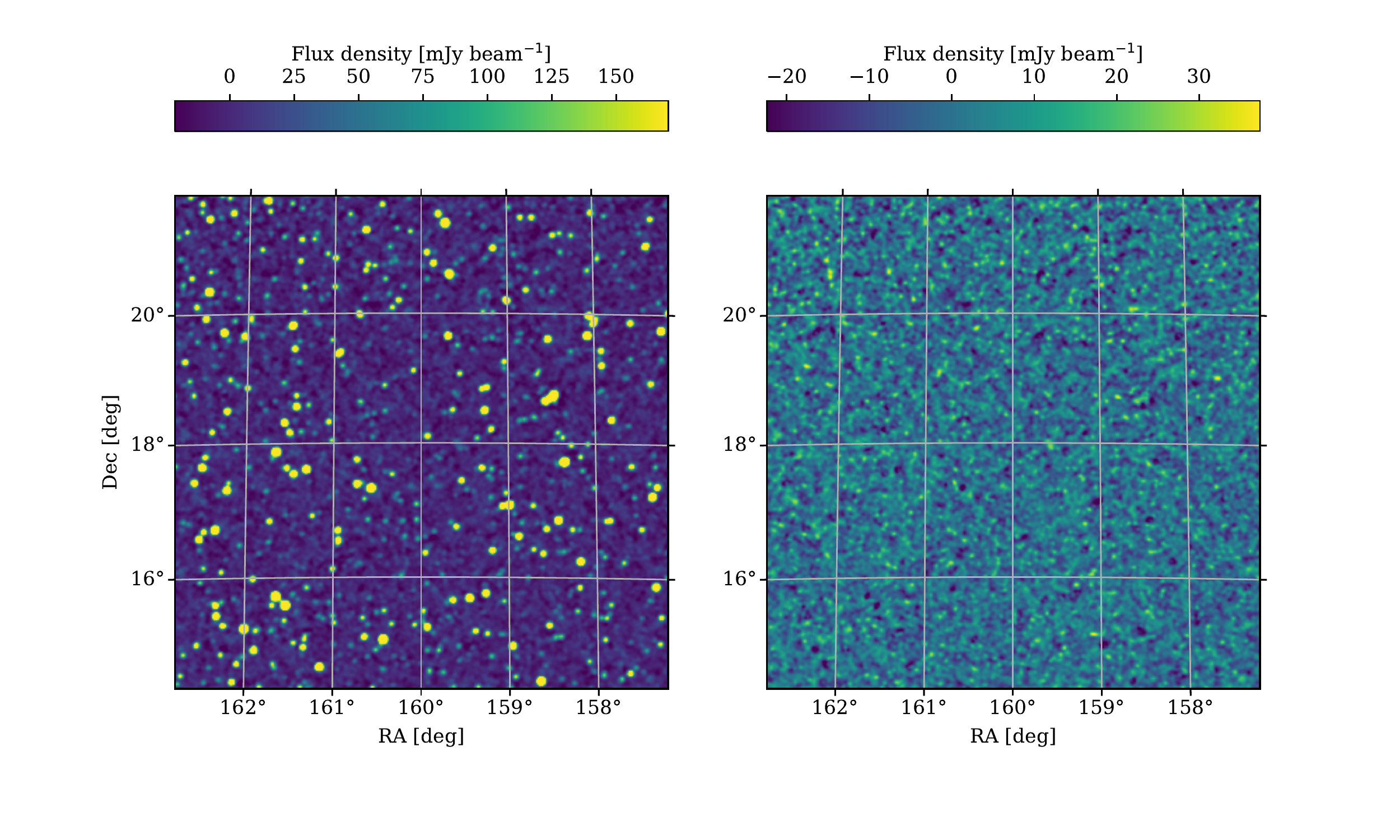}
    \caption{The central region of field 10, centred at RA \SI{160}{\degree}, Dec \SI{18}{\degree}, with \SI{3}{\arcminute} resolution. \textit{Left:} The full mosaic, with all clean components restored into the image. The peak flux density in this image is \SI{3.8}{\jansky \per \beam}, whilst elsewhere in the field it is as high as \SI{34.6}{\jansky \per \beam}. \textit{Right:} The residual image, with all clean components subtracted out. This inset has a mean noise of \SI{7.5}{\milli \jansky \per \beam}, whilst the peak value is \SI{41}{\milli \jansky \per \beam}.}
    \label{fig:field10}
\end{figure*}

In \autoref{fig:field10}, we present a zoom of field 10, showing both the restored and residual images. The brightest source in this field is \SI{34.6}{\jansky \per \beam}. In the residual map, the estimated noise is just \SI{7.5}{\milli \jansky \per \beam}, with a maximum value of \SI{41}{\milli \jansky \per \beam}. The central residual map noise in the other fields is listed in \autoref{tab:fields}.

Prior to stacking, we convert the residual maps from flux density (in units \SI{}{\jansky \per \beam}) to temperature (units \SI{}{\kelvin}), taking into account the spatially varying restoring beam dimensions. In this way, the stacked images are brought to have consistent units despite variable beam sizes, and this is consistent with the method employed in V2021.

\subsection{Exclusion zones}
\label{sec:exclusionzones}

We introduce a number of exclusion zones to our fields to improve the quality of our stacked images. In the first instance, during stacking we truncate around the edges of all fields where the beam power reaches less than 10\% of its peak value. This excludes regions of high noise from being included in our stacks. Then, we visually inspect each field and identify areas to exclude based on two criteria. First, we check for extremely bright sources and draw exclusion zones around them, since their residuals tend to be areas of high noise. In the case of Virgo A, this exclusion zone is sizable in some fields as a result of small calibration errors throwing flux some distance away from the source. Secondly, we search for extended sources that remain in the residuals. Many of these are extended AGN sources that have been cleaned to the level of the noise in individual snapshots but which reappear above the noise once we mosaic, and appear as extended islands of emission typically a few beams in width. These visually inspected regions are collated and nulled in the image prior to stacking.\footnote{For the sake of reproducibility, these exclusion regions are included in the associated data release as DS9 region files.}

\section{Stacking and model subtraction}

Having created deep, well-calibrated images of our 14 fields with point sources subtracted, we can now turn to stacking the LRG pairs. We stack LRG pairs in an effort to drive down the uncorrelated noise in our images, and meanwhile reveal any correlated mean emission that might bridge the LRG pairs. In this section we detail the construction both of these stacked images as well as the process used to construct the LRG models that we ultimately subtract in an effort to detect any excess cosmic web emission.

\subsection{Stacking}

We have implemented our stacking methodology similarly to V2021. We first identify a maximum scaling size, which is at least the maximum pixel distance of any single LRG pair across all fields. All halo pairs are subsequently strictly up-scaled to this size.  We iterate though each LRG pair, once for each field. If the LRG pair is located within the field and does not overlap with an exclusion zone, we proceed to stack this pair. To do this, we identify the pixel coordinates of the pair within the field projection, and calculate both the pixel distance between these coordinates, as well as the angle between their connecting line and horizontal. We rotate and scale the pixel coordinates of the entire field such that pixel distance becomes the maximum pixel distance, and that their connecting line is rotated to horizontal. Finally, we linearly interpolate these values onto a rectangular grid whose centre is the point equidistant the LRG pair. This final map is now ready to be stacked alongside all other LRG pairs.

LRG pairs are weighted by a function of the estimated noise of the field. This estimated noise map is scaled and rotated identically to the field itself. When it comes time to stack, we weight each LRG pair by the inverse square of this map. Note that this noise map is spatially varying and, especially near the edges of the field where the underlying noise is rapidly changing, it is possible for the weighting used for a single LRG pair to vary across the length of the pair. We also track the sum of these weights, and in the final step divide the LRG pair stack by the weight stack to arrive at the weighted mean stack. See \autoref{sec:weighting} for more detail on the stack weighting.

We construct a coordinate system on the final stacked images that places one LRG at $x = -1$, the other at $x = +1$, and the midpoint at the origin. The $y$ direction is scaled identically, and we will herein refer to this as the normalised coordinate system.

\begin{figure}
    \centering
    \vspace{-0.5cm}
    \includegraphics[width=\linewidth,clip,trim={0 0 0 0}]{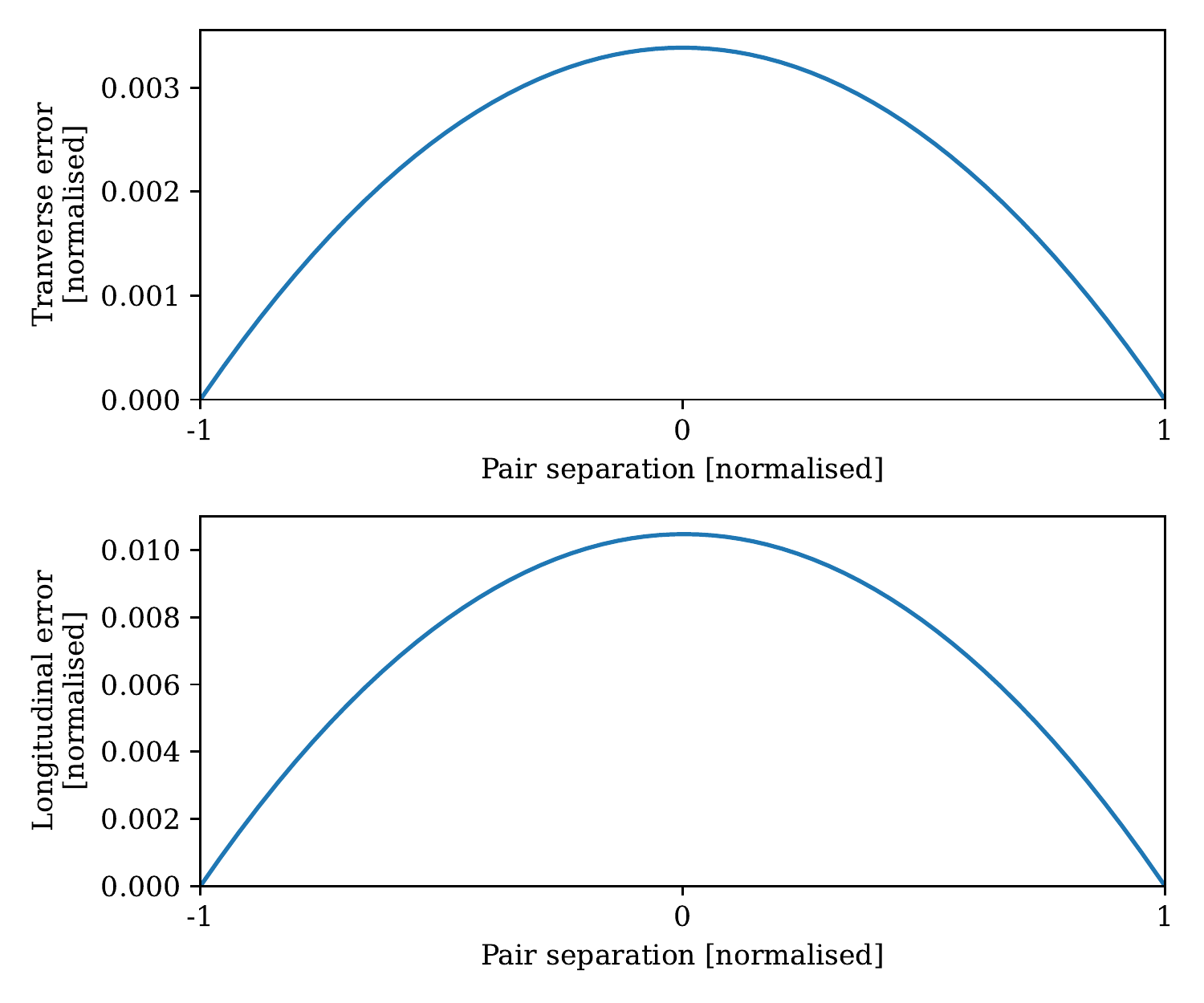}
    \caption{The maximum error associated with treating a SIN projection as a simple Cartesian grid, obtained at the maximum declination \SI{+32}{\degree}. \textit{Top:} The maximum transverse error along a constant-declination \SI{180}{\arcminute} line as a result of geodesics being curved in pixel space. \textit{Bottom:} The maximum longitudinal error along a constant-hour-angle \SI{180}{\arcminute} line, as a result of non-uniform pixel sizes.}
    \label{fig:projectionerror}
\end{figure}

At no point during stacking do we reproject the maps: they are rotated and scaled in pixel coordinates only. An alternative would have been to reproject each pair onto a common projection, but as we have noted earlier, our experience is that such reprojection creates scaling of the flux density values. Using pixel coordinates on an underlying SIN projection, however, has its own downsides whereby: geodesics on the sky are not, in general, straight lines in pixel coordinates; and the angular distance per pixel is not constant. In a SIN projection, these effects are most pronounced at the highest declinations where the field deviates most significantly from a Cartesian grid. They are, however, much smaller than the resolution element of the MWA. For example, in \autoref{fig:projectionerror}, we consider the worst case scenario of an LRG pair at the maximum separation of \SI{180}{\arcminute}, and at the most northern declination of \SI{+32}{\degree}. The upper panel shows the transverse error that results from geodesics not being straight lines in pixel space which peaks at 0.003, whilst the lower panel shows the longitudinal error due to non-uniform pixel sizes which peaks at 0.01. The majority of our LRG pairs have a significantly smaller angular separation, and these errors are markedly smaller in these cases. These errors are small enough that we deem the simplicity and flux correctness of stacking in pixel space to be preferable.

\subsection{Model subtraction}
\label{sec:modelling}

\begin{figure*}
    \centering
    \includegraphics[width=\linewidth,clip,trim={2.9cm 0.5cm 2.7cm 0}]{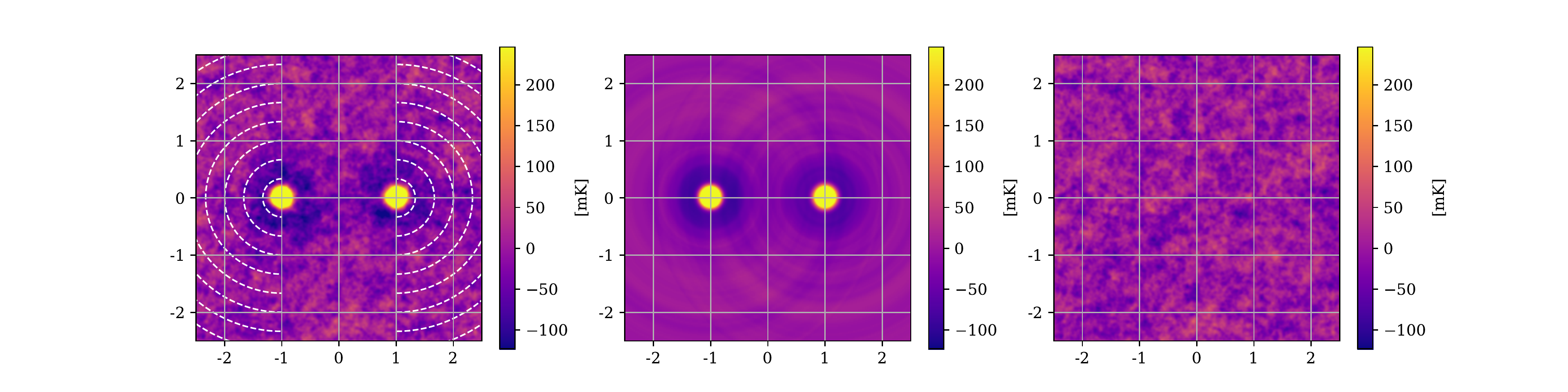}
    \caption{An example showing the LRG model construction and subtraction from the stacked image, with all coordinates in the normalised coordinate system such that the LRG peaks are at $x = \{-1, 1\}$ and the $y$ direction scales identically. \textit{Left:} The original mean stacked image, with the dashed arcs indicating the exterior sweep over which each radially-averaged one-dimensional model is constructed. The LRG peaks rise to just over \SI{4}{\kelvin}; we have set the colour sclae limits on these images to make the noise, at \SI{24.6}{\milli \kelvin}, visible. \textit{Centre:} The model sum map, produced by interpolating the one-dimensional model for each LRG peak onto the two-dimensional map. \textit{Right:} The residual image, after subtracting the model from the original mean stack.}
    \label{fig:examplemodel}
\end{figure*}

Model construction is implemented identically to V2021. It is assumed that emission about each LRG peak, either due to radio emission from the LRG itself or nearby cluster emission, should be radially symmetric. Any cosmic web emission spanning the LRG pair will appear as an excess against this model. Thus, we construct our model based on the \SI{180}{\degree} sweep \textit{exterior} to the LRG pair and we radially average this to form a one-dimensional profile as a function of radial distance. The implementation of this involves binning pixels based on their radial distance, with the bin width set as 1 pixel, before each bin is then averaged. We can then create a function that linearly interpolates over these bins, allowing us to produce a full two dimensional model independently for each LRG as a function of radial distance. Note by creating a model for each LRG peak independently, we are assuming the contribution from each peak is negligible for radial distances $r > 2$. Finally, we sum the LRG model contribution for each peak to produce the final model.

We show an example of this process in \autoref{fig:examplemodel}. In the left panel we show the original mean stacked image. The LRG peaks rise to just over \SI{4}{\kelvin}, however we have set the colour scale limits on these images to make the noise, at \SI{24.6}{\milli \kelvin}, visible. The dashed arcs indicate the exterior sweep over which each radially-averaged one-dimensional model is constructed. These models are then linearly interpolated onto the two-dimensional map and summed, so as to produce the model, shown in the central panel. Finally, we produce the residual stack by subtracting out the model from the mean stack, shown in the rightmost panel. Note the absence of all large-scale structures in the residual, including the LRG peaks themselves as well as the surrounding depressions caused by the MWA dirty beam.

We additionally provide the results of a synthetic test of our stacking and modelling processes in \autoref{sec:synthetic}.

\subsection{Noise characteristics}

\begin{figure}
    \centering
    \includegraphics[width=\linewidth]{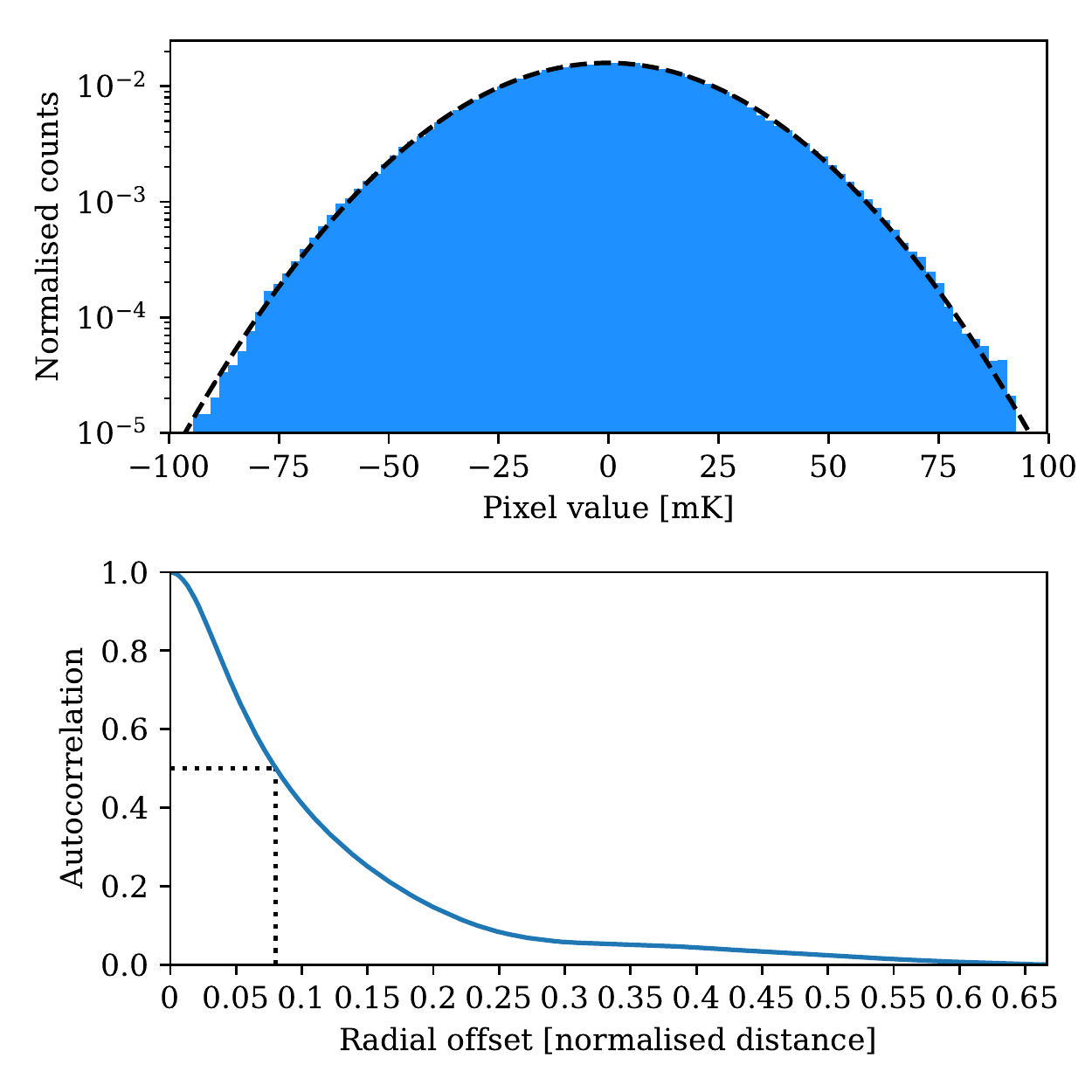}
    \caption{An example of the noise characteristics of the residual stack, in this case from the Max \SI{15}{\mega \parsec} stack. \textit{Top:} The pixel distribution of the residual map, showing an approximately normal distribution. The dashed black line shows the Gaussian fit to the distribution, parameterised as $\sigma =$~\SI{24.6}{\milli \kelvin}. \textit{Bottom:} The radially average autocorrelation of the residual stack, showing the autocorrelation as having a half width at half maximum (dotted black line) of 0.074.}
    \label{fig:noise}
\end{figure}

To determine the significance of any excess signal in the residual stack, it is necessary to characterise the noise of our images. The original fourteen fields consist of real radio emission on top of a background of Gaussian noise. During stacking, the noise in these fields goes down proportionally to the inverse square root of the number of stacks. The presence of real emission peaks in the residual field maps does not affect this, since these peaks are uncorrelated from stack to stack. In the upper panel of \autoref{fig:noise}, we show an example of the pixel distribution of one of our residual stacks, showing that it very nearly approximates a normal distribution, as indicated by the dashed black line, with $\sigma =$~\SI{24.6}{\milli \kelvin}. All our stacks exhibit this kind of normal distribution of pixel values, and so we will characterise them by reference to the standard deviation of their residual maps.

The noise, however, is spatially correlated. In the original fields prior to stacking, this spatial correlation is on the scale of dirty beam. In the stacked images, however, this is not the case, since during stacking we rescale each LRG pair. To characterise the effective resolution of the stacked image we perform a radially averaged two-dimensional auto-correlation of the residual stack, and we present an example of this in the lower panel of \autoref{fig:noise}. We observe in this plot both an extended peak in this function, showing the spatial correlation of pixels persists in the stacked images, and also a slight depression showing the cumulative sidelobes of the stacked, dirty beams. We characterise the effective resolution by measuring the full width at half maximum (FWHM). In this case, the half width at half maximum of the autocorrelation is 0.074, corresponding to a FWHM value of the residual map of 0.105.

These two metrics---the standard deviation and the effective resolution---allow us to understand the significance of any potential signal in our stacks. Specifically, peaks of excess emission that deviate significantly from the measured map noise, or extended emission on scales greater than the effective resolution, are tell-tale markers that we are encountering signal that deviates from otherwise stochastic noise.

\section{Results and Discussion}

\subsection{Stacking results}

\begin{figure*}
    \centering
    \begin{subfigure}{\linewidth}
        \centering
        \includegraphics[width=0.7\linewidth,clip,trim={2cm 1cm 0cm 2cm}]{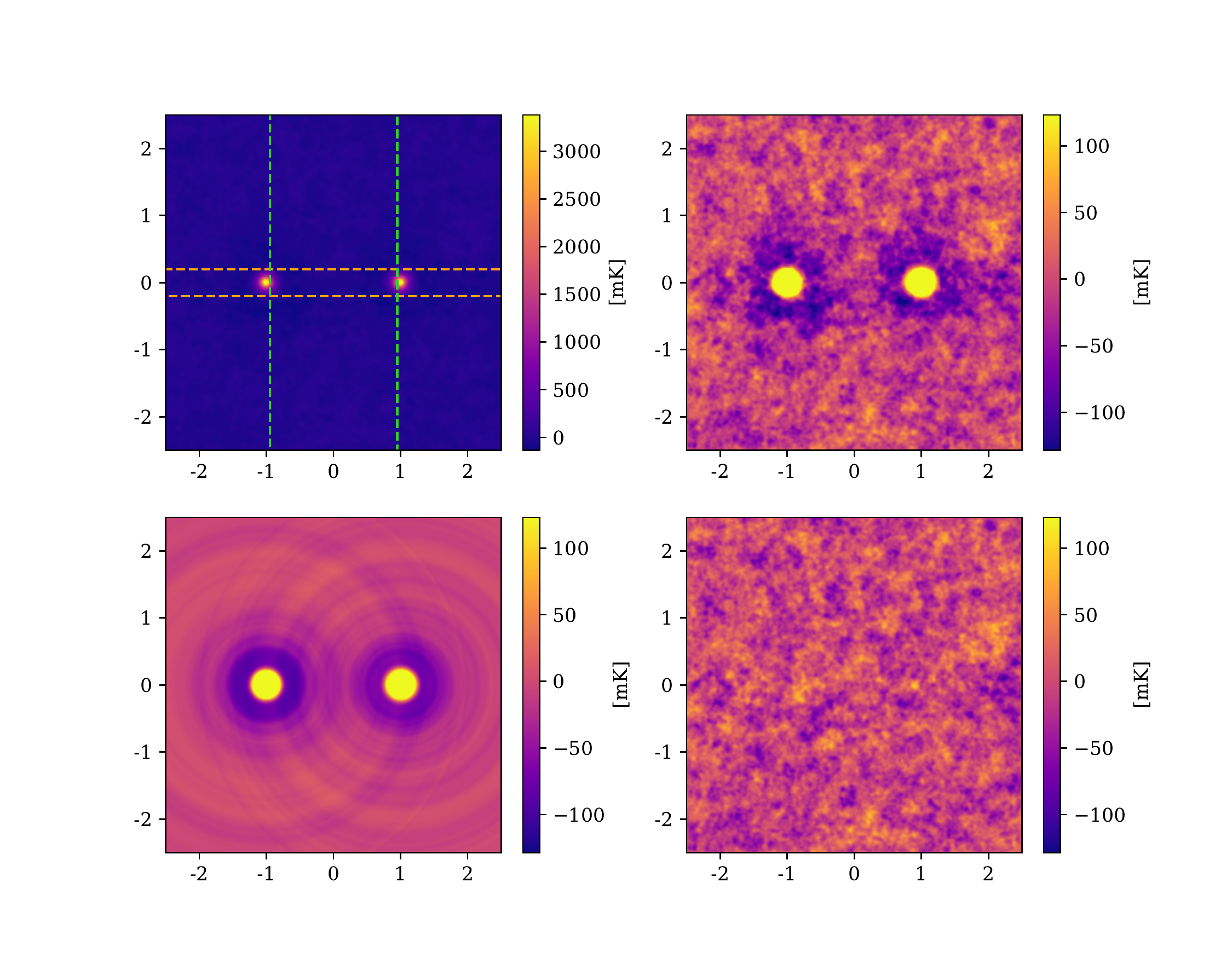}
        \caption{\textit{Top left:} The original mean stack image, with overlays indicating the region over which the transverse mean (dashed orange horizontal lines) and longitudinal mean (dashed green vertical lines) are calculated. \textit{Top right:} The mean stacked image with the colour scale reduced to $\pm 5 \sigma$ to emphasise the noise. \textit{Bottom left:} The model image, on the same colour scale. \textit{Bottom right:} The residual stack after model subtraction, with the colour scale set to $\pm 5 \sigma$.} 
        \label{fig:max15a}
    \end{subfigure}
    \begin{subfigure}{\linewidth}
        \centering
        \includegraphics[width=0.8\linewidth,clip,trim={0 0 0 -0.5cm}]{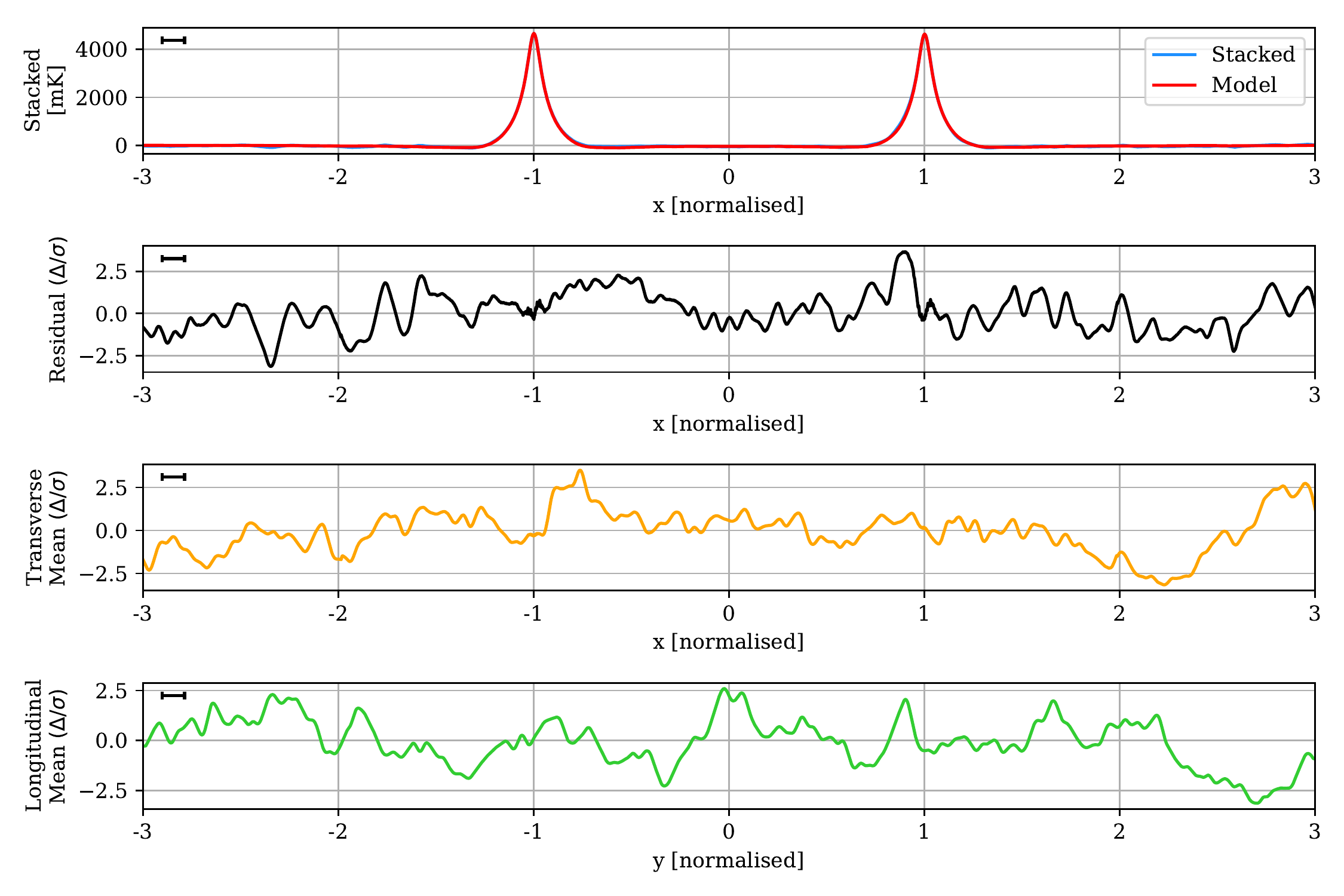}
        \caption{\textit{One:} The one-dimensional profile along $y = 0$ for both the stacked image (blue) and the model (red). \textit{Two:} The one-dimensional profile along $y = 0$ of the residual stack, renormalised to the estimated map noise. \textit{Three:} The transverse mean along the region $-0.2 < y < 0.2$ of the residual stack, renormalised to the estimated map noise. \textit{Four:} The longitudinal mean along the region $-0.95 < x < 0.95$ of the residual stack, renormalised to the estimated map noise. The black rule in the top left shows the FHWM of the effective resolution.}
        \label{fig:max15b}
    \end{subfigure}
    \caption{The Max \SI{15}{\mega \parsec} stack, with mean LRG peaks of \SI{4292}{\milli \kelvin}, residual noise of \SI{25}{\milli \kelvin}, and effective resolution of 0.11.}
    \label{fig:max15}
\end{figure*}

\begin{figure*}
    \centering
    \begin{subfigure}{\linewidth}
        \centering
        \includegraphics[width=0.7\linewidth,clip,trim={2cm 1cm 0cm 2cm}]{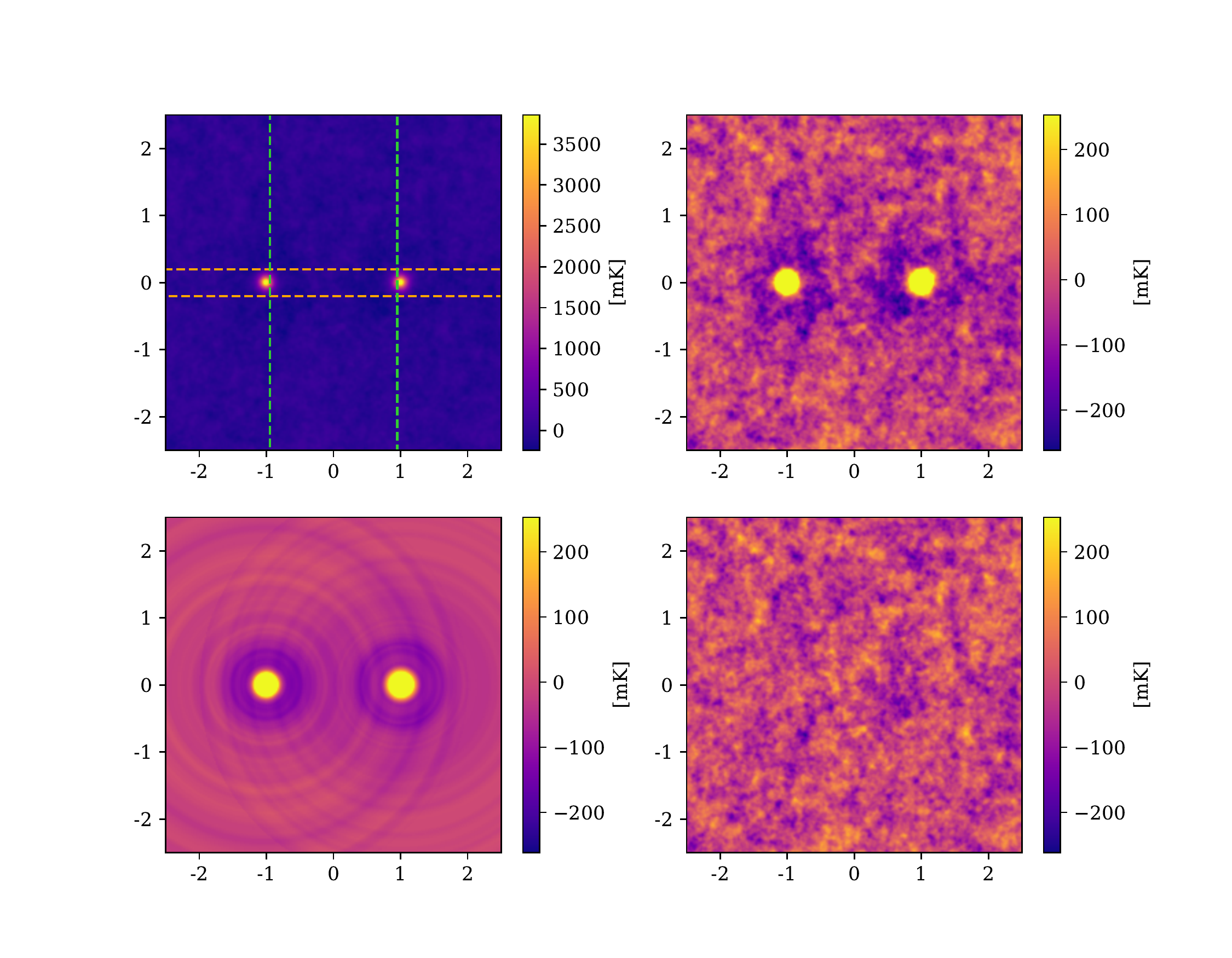}
        \caption{\textit{Top left:} The original mean stack image, with overlays indicating the region over which the transverse mean (dashed orange horizontal lines) and longitudinal mean (dashed green vertical lines) are calculated. \textit{Top right:} The mean stacked image with the colour scale reduced to $\pm 5 \sigma$ to emphasise the noise. \textit{Bottom left:} The model image, on the same colour scale. \textit{Bottom right:} The residual stack after model subtraction, with the colour scale set to $\pm 5 \sigma$.} 
    \end{subfigure}
    \begin{subfigure}{\linewidth}
        \centering
        \includegraphics[width=0.8\linewidth,clip,trim={0 0 0 -0.5cm}]{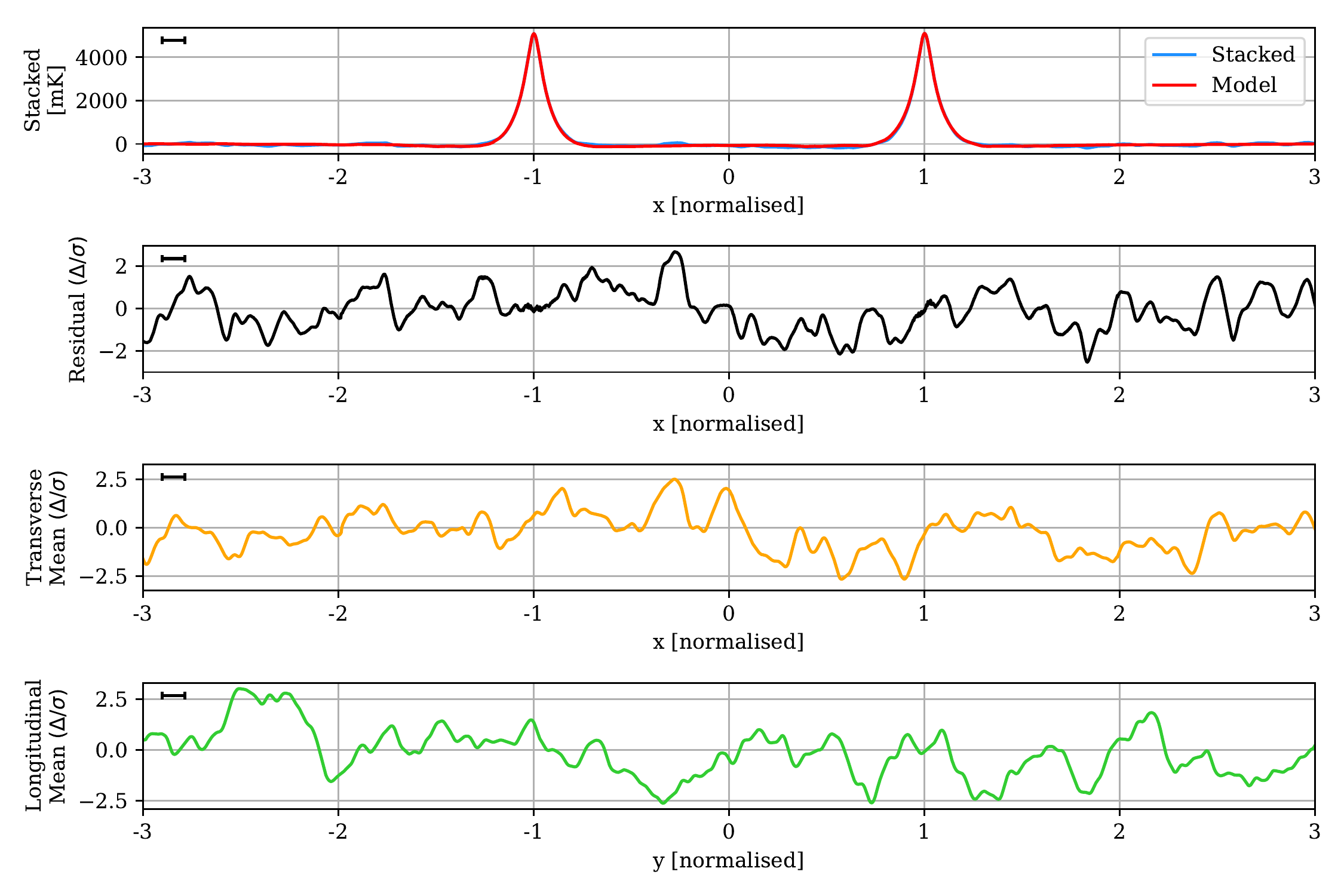}
        \caption{\textit{One:} The one-dimensional profile along $y = 0$ for both the stacked image (blue) and the model (red). \textit{Two:} The one-dimensional profile along $y = 0$ of the residual stack, renormalised to the estimated map noise. \textit{Three:} The transverse mean along the region $-0.2 < y < 0.2$ of the residual stack, renormalised to the estimated map noise. \textit{Four:} The longitudinal mean along the region $-0.95 < x < 0.95$ of the residual stack, renormalised to the estimated map noise. The black rule in the top left shows the FHWM of the effective resolution.}
    \end{subfigure}
    \caption{The Max \SI{10}{\mega \parsec} stack, with mean LRG peaks of \SI{4699}{\milli \kelvin}, residual noise of \SI{51}{\milli \kelvin}, and effective resolution of 0.12.}
    \label{fig:max10}
\end{figure*}

\begin{figure*}
    \centering
    \begin{subfigure}{\linewidth}
        \centering
        \includegraphics[width=0.7\linewidth,clip,trim={2cm 1cm 0cm 2cm}]{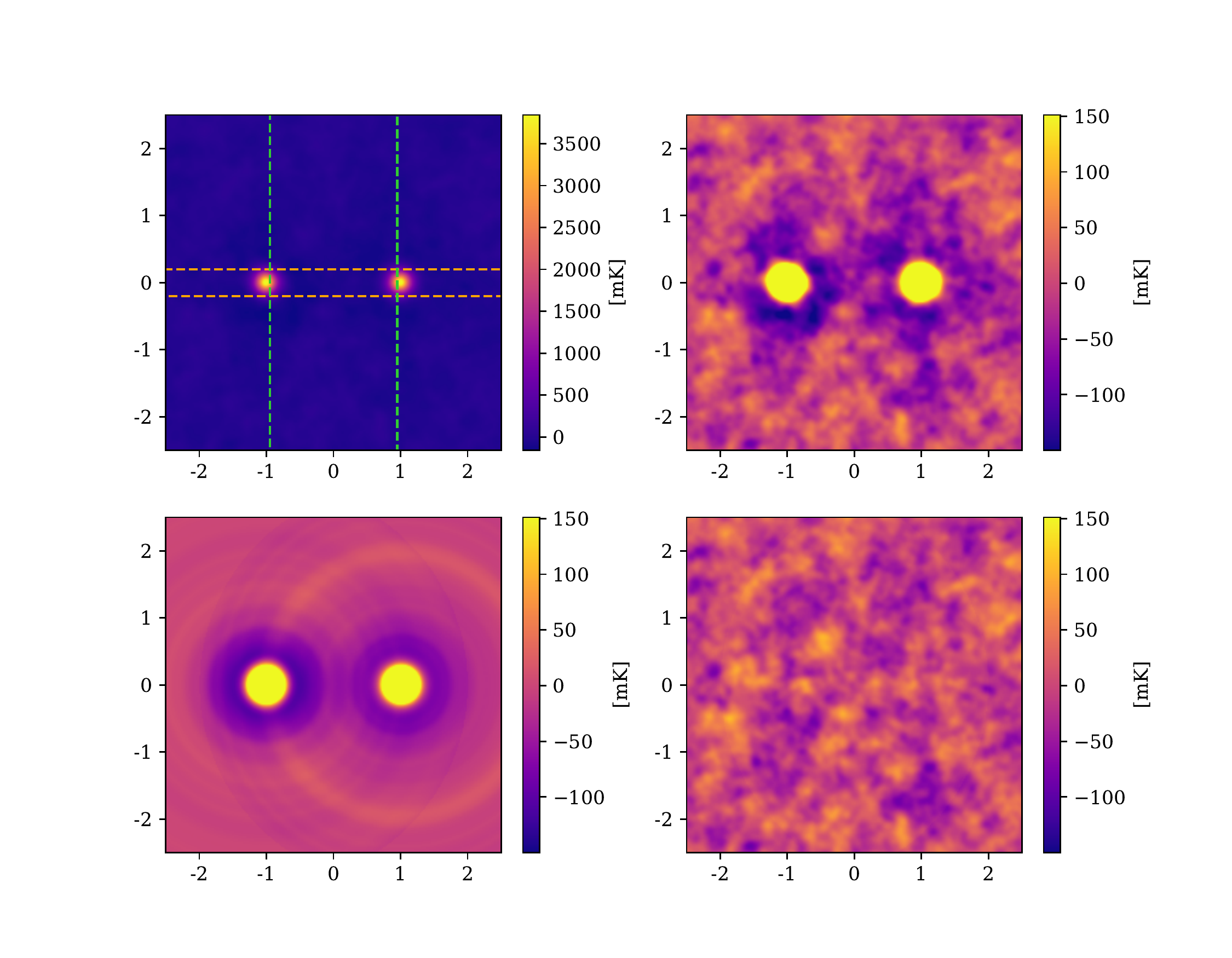}
        \caption{\textit{Top left:} The original mean stack image, with overlays indicating the region over which the transverse mean (dashed orange horizontal lines) and longitudinal mean (dashed green vertical lines) are calculated. \textit{Top right:} The mean stacked image with the colour scale reduced to $\pm 5 \sigma$ to emphasise the noise. \textit{Bottom left:} The model image, on the same colour scale. \textit{Bottom right:} The residual stack after model subtraction, with the colour scale set to $\pm 5 \sigma$.} 
    \end{subfigure}
    \begin{subfigure}{\linewidth}
        \centering
        \includegraphics[width=0.8\linewidth,clip,trim={0 0 0 -0.5cm}]{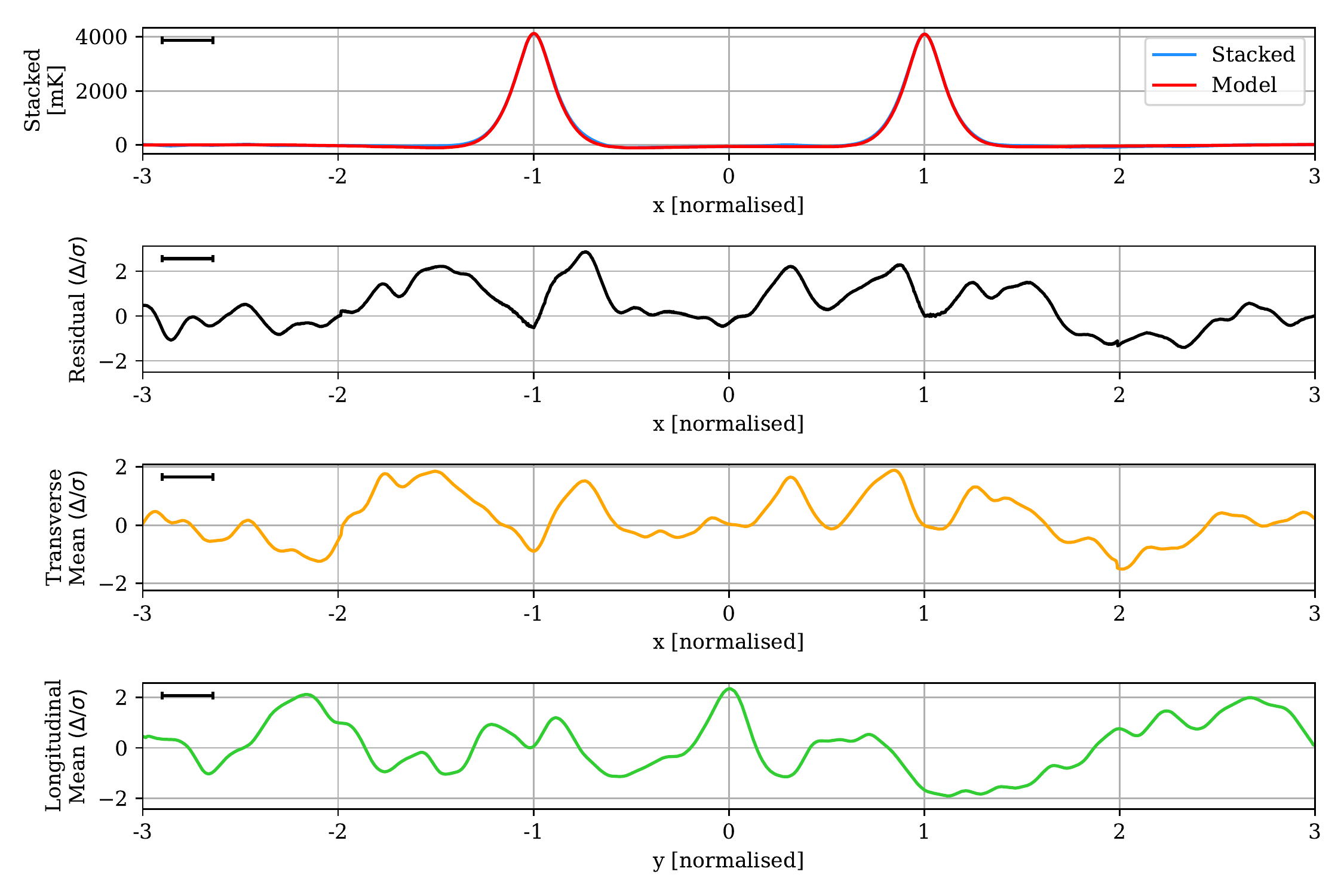}
        \caption{\textit{One:} The one-dimensional profile along $y = 0$ for both the stacked image (blue) and the model (red). \textit{Two:} The one-dimensional profile along $y = 0$ of the residual stack, renormalised to the estimated map noise. \textit{Three:} The transverse mean along the region $-0.2 < y < 0.2$ of the residual stack, renormalised to the estimated map noise. \textit{Four:} The longitudinal mean along the region $-0.95 < x < 0.95$ of the residual stack, renormalised to the estimated map noise. The black rule in the top left shows the FHWM of the effective resolution.}
    \end{subfigure}
    \caption{The Max \SI{60}{\arcminute} stack, with mean LRG peaks of \SI{3769}{\milli \kelvin}, residual noise of \SI{30}{\milli \kelvin}, and effective resolution of 0.26.}
    \label{fig:max60}
\end{figure*}

\begin{figure*}
    \centering
    \begin{subfigure}{\linewidth}
        \centering
        \includegraphics[width=0.7\linewidth,clip,trim={2cm 1cm 0cm 2cm}]{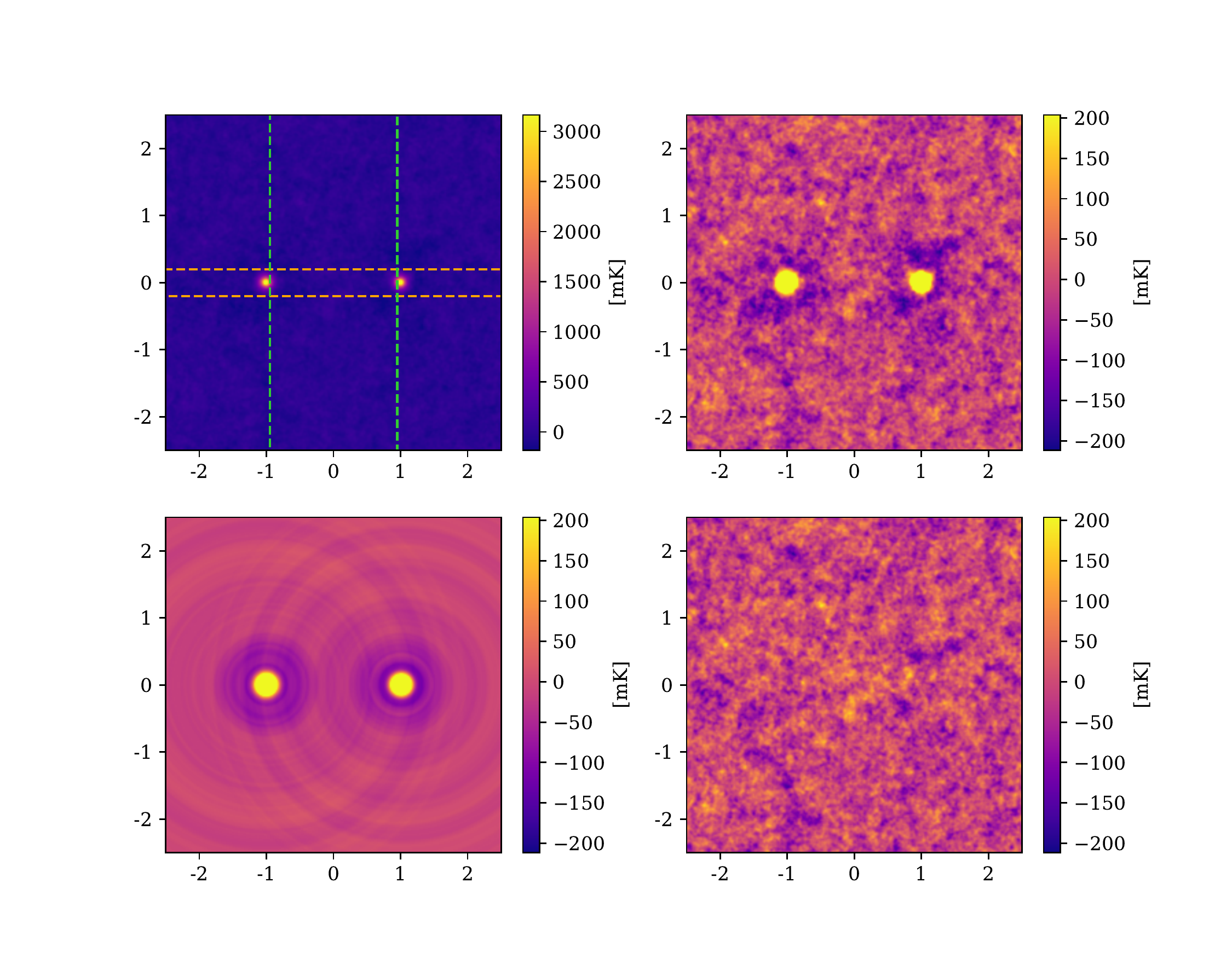}
        \caption{\textit{Top left:} The original mean stack image, with overlays indicating the region over which the transverse mean (dashed orange horizontal lines) and longitudinal mean (dashed green vertical lines) are calculated. \textit{Top right:} The mean stacked image with the colour scale reduced to $\pm 5 \sigma$ to emphasise the noise. \textit{Bottom left:} The model image, on the same colour scale. \textit{Bottom right:} The residual stack after model subtraction, with the colour scale set to $\pm 5 \sigma$.} 
    \end{subfigure}
    \begin{subfigure}{\linewidth}
        \centering
        \includegraphics[width=0.8\linewidth,clip,trim={0 0 0 -0.5cm}]{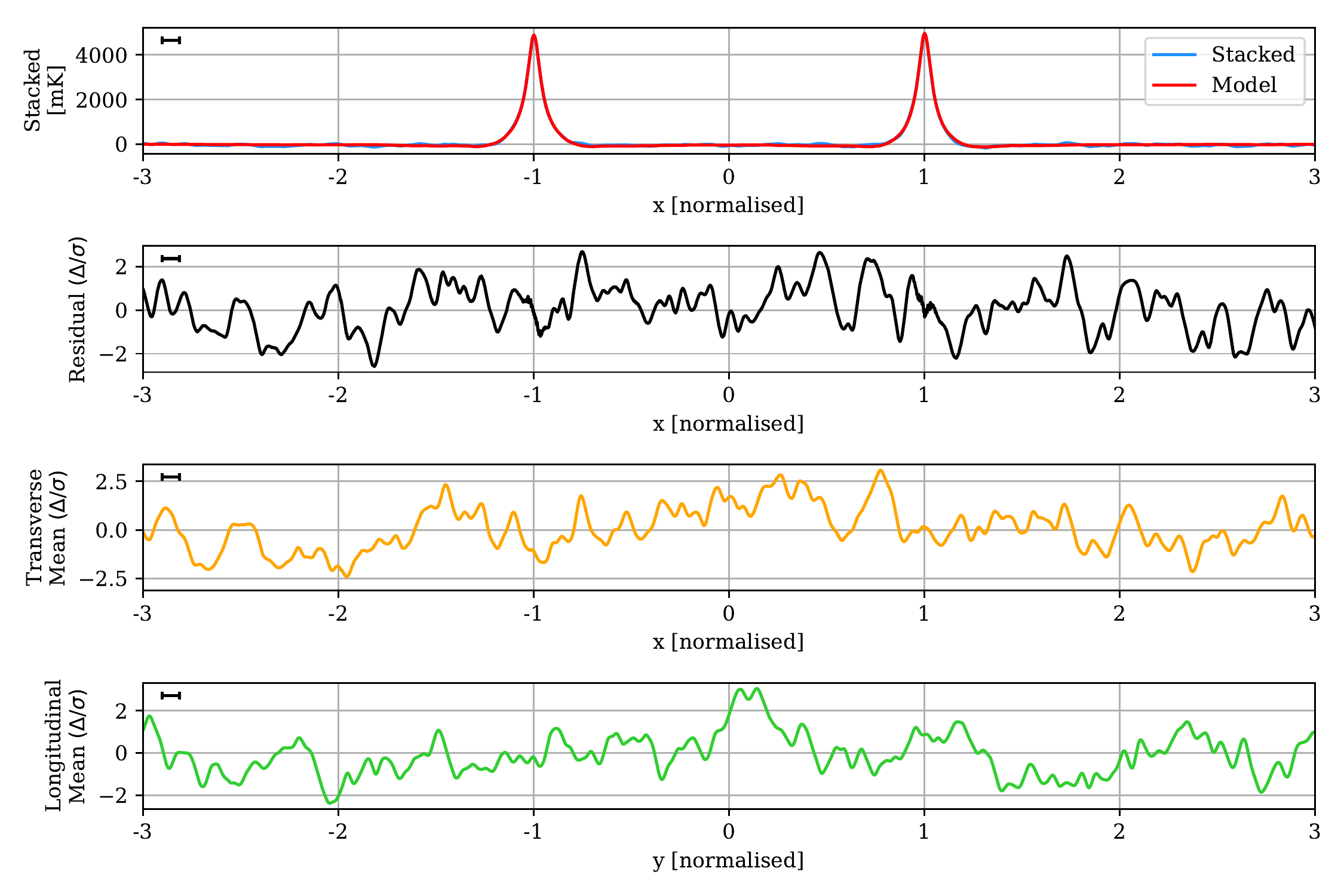}
        \caption{\textit{One:} The one-dimensional profile along $y = 0$ for both the stacked image (blue) and the model (red). \textit{Two:} The one-dimensional profile along $y = 0$ of the residual stack, renormalised to the estimated map noise. \textit{Three:} The transverse mean along the region $-0.2 < y < 0.2$ of the residual stack, renormalised to the estimated map noise. \textit{Four:} The longitudinal mean along the region $-0.95 < x < 0.95$ of the residual stack, renormalised to the estimated map noise. The black rule in the top left shows the FHWM of the effective resolution.}
    \end{subfigure}
    \caption{The LRG-V2021 stack, with mean LRG peaks of \SI{4540}{\milli \kelvin}, residual noise of \SI{42}{\milli \kelvin}, and effective resolution of 0.09.}
    \label{fig:v2021}
\end{figure*}

In \autoref{fig:max15}, we present the stacked results for the Max \SI{15}{\mega \parsec} catalogue. This catalogue consists of 601,435 LRG pairs, allowing our stack to reach a noise of \SI{25}{\milli \kelvin}, more than twice as deep as the \SI{118.5}{\mega \hertz} stack in V2021. As can be seen in the upper left panel of \autoref{fig:max15a}, the peaks at $x = \pm 1$ are the dominant features, and have a mean value of \SI{4292}{\milli \kelvin}. The upper right panel in \autoref{fig:max15a} shows the same stacked image, only with the colour scale adjusted down so as to emphasise the noise. We now note the shallow depressions around each of the peaks, which are attributable to the dirty beam's negative sidelobes. The LRG model is shown in the bottom left panel, and the bottom right panel in \autoref{fig:max15a} shows the residual stack, after model subtraction. The model construction methodology is surprisingly effective, leaving no trace either of the sharp peaks at $x = \pm 1$, as well as removing the sidelobe depressions. There is no readily apparent excess emission in the residual image. In the top panel of \autoref{fig:max15b}, we compare the one-dimensional slice through $y = 0$ of both the mean stacked image (blue) and model (red). The stacked image and model are so similar that we scarcely observe any of the stacked plot. Note that the widths of the peaks are narrower than observed in V2021: the peaks here have a FWHM value of 0.11, and whilst this value is not given in V2021, their peaks appear visually much wider. These peaks will be in part a function of the instrumental dirty beam, however this is not sufficient to explain this discrepancy; we discuss this more in \autoref{sec:peakwidth}. In the second panel of \autoref{fig:max15b}, we show the one-dimensional $y = 0$ slice through the residual image, where we have renormalised the scale to the estimated map noise. There are no peaks in this residual exceeding $3 \sigma$. In the third panel, we display the mean value in the range $y = \pm 0.2$ as a function of $x$, and renormalise based on the estimated map noise. The aim of this transverse mean is to bring out faint, wide signals that might be present along the intercluster stacks. For this LRG catalogue, we observe no peaks exceeding $3 \sigma$. Finally, in the lower panel we display the longitudinal mean in the range $-0.95 < x < 0.95$, as a function of $y$. For a faint signal that spans the length of the intercluster stack, we would expect this plot to show a peak at $y = 0$, however we observe no statistically significant signal. We conclude there is no statistically significant excess emission along the bridge for the Max \SI{15}{\mega \parsec} stack.

The stacked results for the Max \SI{10}{\mega \parsec} catalogue are shown in \autoref{fig:max10}. With just a quarter of the LRG pairs as the larger Max \SI{15}{\mega \parsec} catalogue, the estimated noise of this stack is higher at \SI{51}{\milli \kelvin}, just a slight improvement on the stated noise in the \SI{118.5}{\mega \hertz} stack in V2021. The peaks at $x = \pm 1$ are higher than the previous stacks, at \SI{4699}{\milli \kelvin}, which is a result of the catalogue sampling from a more local redshift space, whilst their widths have a similar FWHM of 0.12. Once again, however, the residual image and one-dimensional slices show no indication of statistically significant excess emission along the bridge.

Likewise, the stacked results for the Max \SI{60}{\arcminute} and LRG-V2021 catalogues, in \autoref{fig:max60} and \autoref{fig:v2021} respectively, also show no evidence of excess emission. The Max \SI{60}{\arcminute} stack has a noise of \SI{30}{\milli \kelvin} and a large effective resolution of 0.26 that is a result of reduced lower angular threshold and corresponding variation in scaling during stacking. Similarly, the peak width has increased to 0.23. This LRG catalogue also samples significantly deeper in redshift space than the others, with the result that the LRG peaks are diminished in comparison, with a mean value of \SI{3769}{\milli \kelvin}. One small $\sim2.85\sigma$ peak is visible in the one-dimensional profile at $x = -0.73$, however its width matches the effective resolution, and similar peaks throughout the residual image suggest it is consistent with the noise. Meanwhile, the LRG-V2021 stack has a noise of \SI{41}{\milli \kelvin}, approximately \SI{30}{\percent} lower than the equivalent \SI{118.5}{\mega \hertz} stack in V2021. It has a peak in the longitudinal profile at $y = 0.14$ that reaches a significance of $3.04 \sigma$, but otherwise shows no evidence of intercluster signal and certainly not the kind of large-scale, clearly evident excess emission as shown in V2021.

The analysis of each of our LRG catalogue stacks leaves us unable to corroborate the detection of V2021.

\subsection{Sensitivity to extended emission}
\label{sec:resolvingout}

\begin{figure*}
    \centering
    \includegraphics[width=\linewidth,clip,trim={4cm 0 4cm 0}]{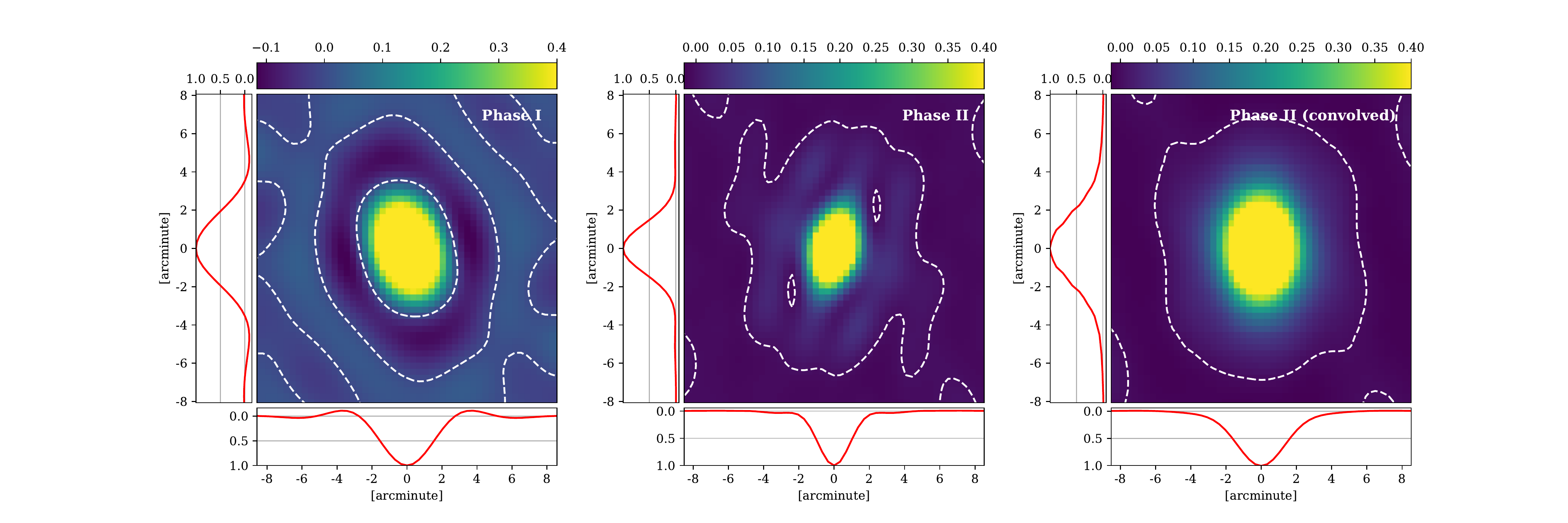}
    \caption{A comparison of dirty beams used in V2021 and the present study, measured at \SI{118.5}{\mega \hertz} and pointing $\alpha =$~\SI{180}{\degree} $\delta =$~\SI{18}{\degree}. {White dashed contours trace a response of zero, so as to better show the negative sidelobe regions.} \textit{Left:} The Phase I dirty beam with baseline weighting Briggs -1, as used in GLEAM, having a resolution of \SI{3.74 x 2.56}{\arcminute}. \textit{Centre:} The Phase II dirty beam with baseline weighting Briggs +1, as used in the current study, and having a resolution of \SI{3.2 x 1.9}{\arcminute}. \textit{Right:} The Phase II dirty beam, after convolution, having a resolution of \SI{4.2 x 3.1}{\arcminute}.}
    \label{fig:psfs}
\end{figure*}

\begin{figure}
    \centering
    \includegraphics[width=\linewidth]{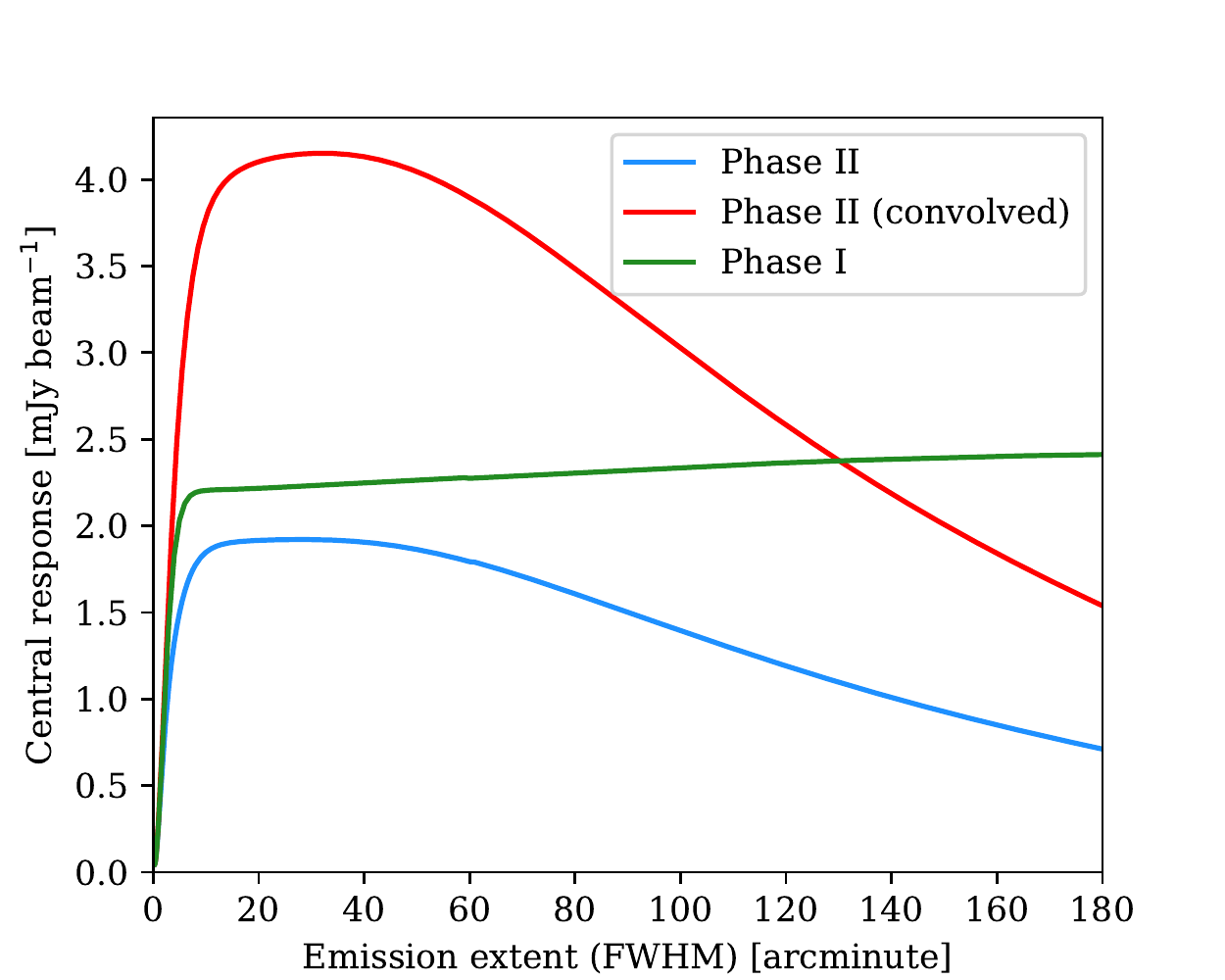}
    \caption{The sensitivity of Phase I, II, and Phase II (convolved) to extended emission. The plot shows the response at the centre of simulated circular Gaussians of varying sizes, with the simulated sources having a constant peak surface brightness of \SI{1}{\jansky \per \square \deg}. For large, extended emission sources, there exists a threshold angular scale above which the central response begins to drop, as these sources become increasingly `resolved out'. On the other hand, for very small angular sizes, the simulated source becomes smaller than the dirty beam (i.e.\@ is unresolved) whilst maintaining the same peak surface brightness; the total flux of the source thus rapidly drops to zero as does the instrumental response.}
    \label{fig:sensitivity}
\end{figure}

The chief distinction between the MWA Phase I and Phase II instruments is the location of the antennas, and in turn, each instrument's respective dirty beam. As noted previously, the point source sensitivity is unchanged. However, these modified baselines may make the instrument less sensitive to extended emission, potentially even resolving out large-scale emission such as the cosmic web, and this may be a factor in our non-detection.

In \autoref{fig:psfs} we show the dirty beams of the Phase I and Phase II instrument, as well as the effective dirty beam of the Phase II instrument after our convolution to \SI{3}{\arcminute} (at zenith) resolution. These dirty beams have been generated from archival \SI{118.5}{\mega \hertz} MWA observations at the centre of field 11 ($\alpha =$~\SI{180}{\degree}, $\delta =$~\SI{18}{\degree}) to best model the effect of the low elevation pointings on the dirty beam. The Phase I dirty beam is produced with a Briggs -1 baseline weighting scheme such that it matches the original GLEAM imaging parameters, and has a resolution of approximately \SI{3.74 x 2.56}{\arcminute}. Note the sizeable negative sidelobes around the beam, owing to a dense core of short baselines. The Phase II dirty beam is produced with the same baseline weighting as used in the present work, Briggs +1, as well as its lower baseline length threshold of $15 \lambda$, and has a resolution of approximately \SI{3.2 x 1.9}{\arcminute}. After convolution, this grows to a resolution of \SI{4.2 x 3.1}{\arcminute} at the centre of the field.

% Radio astronomers have often used a back-of-the-envelope method to calculate sensitivity to large scale emission, where sensitivity to emission on angular scales $\theta$ depends on the the length of the shortest baseline $d$ and the observing wavelength $\lambda$, given as $\theta = \nicefrac{\lambda}{d}$. This would suggest the MWA Phase II, with a baseline cut at $15 \lambda$, should be sensitive on scals of XXX. Whilst this formalism accurately measures the angular separation of individual baseline fringes, it does not properly account for the more complex effects of the superposition of all of an array's baselines, which ultimately form its dirty beam. In practice, these effects result in an instrument beginning to resolve out large scale emission on much smaller scales.

\citet{Hodgson2020} developed an empirical method to measure an instrument's sensitivity to large-scale emission, which we draw on here. Often angular sensitivity is estimated solely based on the angular size of the fringe patterns of the shortest baselines in an array, however, this does not take into account the imaging parameters, baseline weightings, and most importantly, the cumulative effect of the instrument's baselines in determining angular sensitivity. Instead, the method we use here proceeds by simulating a range of extended emission sources---in our present case circular Gaussian sources---directly into the visibilities of an observation, and then producing a dirty image of the source. Given a surface flux density of \SI{1}{\jansky \per \square \deg} at the Gaussian peak, we can understand the instrument's response by measuring the flux density at the centre of the Gaussian in the dirty image. If we iterate through many such circular Gaussian sources of increasing size, we will identify a threshold angular scale at which point the central response will begin to reduce, above which scales the dirty image of the Gaussian will start to `hollow out' in the centre and become increasingly dark. In this way we can identify the relative sensitivity of the instrument over a range of angular scales as well as the angular scale at which emission begins to resolve out.

In \autoref{fig:sensitivity} we show the results of this exercise, where we have measured the central response to circular Gaussians having a FWHM up to \SI{180}{\arcminute} in extent. It is immediately apparent that the larger beam size of the Phase I instrument makes it more sensitive than Phase II to large scale emission features, as we'd expect. Moreover, the Phase I instrument does not begin to resolve out structure on these spatial scales; in fact, it continues to gain sensitivity over this range. The sensitivity of the Phase II instrument, on the other hand, begins to slowly decline on angular scales larger than \SI{30}{\arcminute}, and then more rapidly decline on scales larger than approximately \SI{50}{\arcminute}. The effect of convolving the Phase II dirty beam is dramatic, amplifying its sensitivity to extended sources more than a factor of two. Crucially, it also makes the instrument more sensitive than Phase I. It does not forestall the angular scales on which the instrument begins to resolve out structure, however it remains more sensitive than Phase I to extended emission out to approximately \SI{130}{\arcminute}.

When considering whether these differences in sensitivity to extended emission can account for our non-detection of the synchrotron cosmic web we need to understand the typical angular scales we would expect. In the first instance, the majority of LRG pairs in each catalogue are separated by less than \SI{60}{\arcminute}, with the exception of the LRG-V2021 catalogue which has a median separation of \SI{79}{\arcminute}. We should expect our observations to be at least as sensitive as V2021 for those LRG pairs with separations less than \SI{60}{\arcminute}, and specifically with regards to the Max \SI{60}{\arcminute} stack, there is no risk of resolving out structure across the entirety of its LRG pair catalogue. Secondly, we do not expect the emission spanning the intercluster region to be as wide as it is long: whilst our selection criteria allows for these bridges to span up to \SI{180}{\arcminute}, we should expect the width of the bridge to be significantly more narrow. Any MWA baseline fringes aligned approximately along the narrower width will not be at risk of resolving out the emission, and this will reduce the overall effect. Finally, as simulations by \citet{Vazza2019} and further showcased in \citet{Hodgson2021a} have shown, the morphology of the cosmic web is expected to consist of radio relic-like accretion shocks. These typically appear as long extended arcs of emission, usually with a well-defined edge along the shock itself, with many such shocks spanning the length of the intercluster region. Crucially, these kinds of emission mechanisms do not form a broad, continuous bridge of emission that we might risk resolving out, rather they are punctuated and individually consist of sharply defined edges that interferometers are well-suited to detect.

For these reasons, we do not believe we are adversely affected by the higher resolution MWA Phase II instrument.

\subsection{The expected peak widths}
\label{sec:peakwidth}

\begin{figure}
    \centering
    \includegraphics[width=\linewidth]{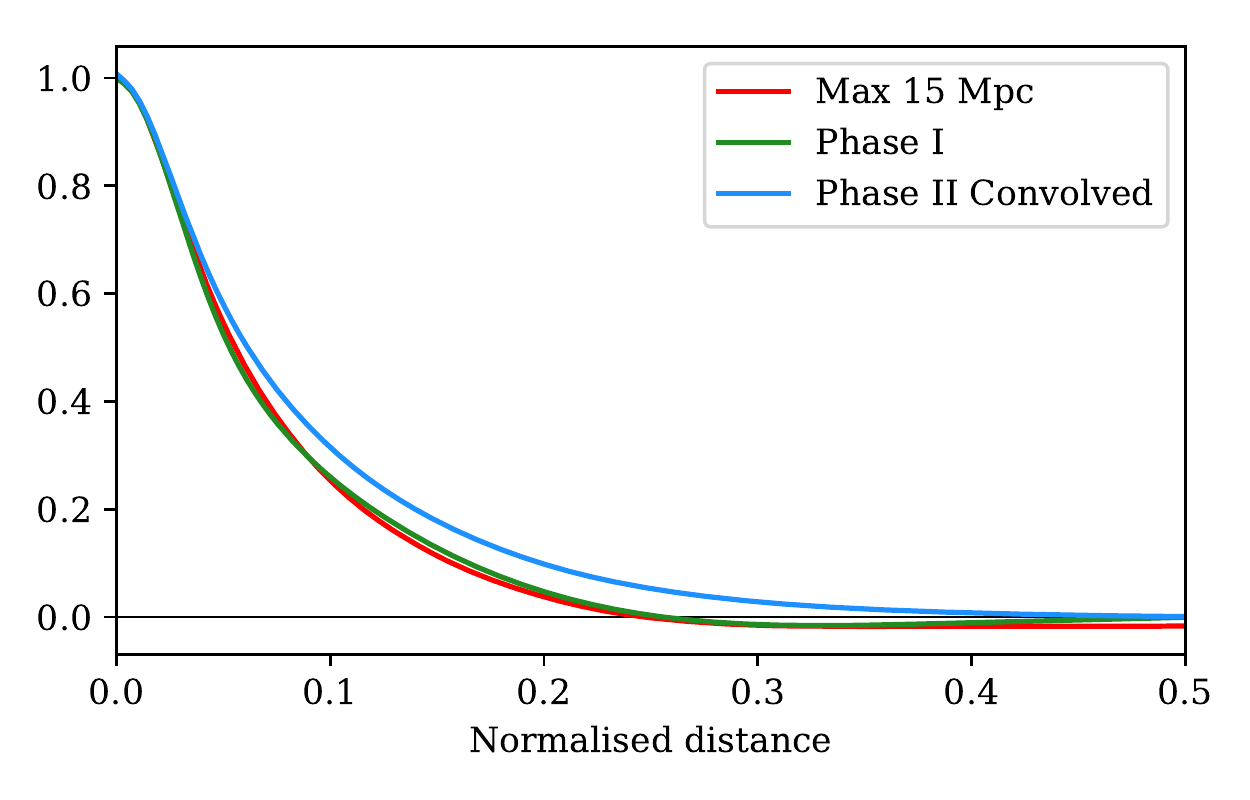}
    \caption{One dimensional profiles of stacked dirty beams for Phase I (green) and Phase II (convolved; blue), in comparison to the Max 15 Mpc stacked model profile (red). The stacked dirty beams approximate a minimum peak profile for purely unresolved LRG sources, and the similarity to the Max 15 Mpc stacked model profile suggests this profile is dominated principally by unresolved sources.}
    \label{fig:psfstack}
\end{figure}

 As noted, a key difference between our results and those of V2021 is that the width of the peaks at $x = \{-1, 1\}$ of our stacks are much narrower. In this section, we want to understand the expected minimum size of these peaks. This condition of minimum peak width occurs when the angular scale of the LRG emission (or other spatially correlated emission) is much smaller than the instrumental dirty beam, that is, when the LRG emission is unresolved and approximately `point-like'. In this case, the instrumental response is simply the dirty beam itself. We can then model the expected minimum-width LRG peak profile by stacking the dirty beam in the following way: for each LRG pair, we calculate the angular distance between the pair and find a scaling factor to upscale onto the maximum angular separation, being \SI{3}{\degree}; we use this scaling factor to stretch the one dimensional profile of the dirty beam; and then sum this alongside other similarly scaled profiles. We build in two additional assumptions in this simple model: first, for each pair we create a one-dimensional profile of the dirty beam at a uniformly random angle through the two-dimensional peak response, which assumes that the orientation of LRG pairs on the sky are approximately uniform; second, that each LRG has an equal contribution to the sum. A key limitation of this exercise is the use of a single dirty beam, as shown previously in \autoref{fig:psfs}; these dirty beams have been generated for a fixed position on the sky, and at these low elevations the dirty beam is especially sensitive to the foreshortening effects of declination changes. Nonetheless, this exercise will give us a good approximation of the minimum peak sizes.
 
 We show the results of this exercise in \autoref{fig:psfstack} for both the Phase I (green) and Phase II (convolved; blue) dirty beams calculated across the Max 15 Mpc catalogue, as well as the model profile of the left peak of the Max 15 Mpc stack (red), shown previously in \autoref{fig:max15}. The FWHM of the Phase II (convolved) peak is 0.12, which compares to the actual model peak width of 0.11. The similarity in both the peak shape and width between this exercise and the actual model suggests the peaks in our stacks are dominated principally by unresolved sources.
 
 In comparison, the peak widths of the stacks in V2021 appear significantly wider. In \autoref{fig:psfstack}, we also show the results of the same exercise for the Phase I dirty beam, showing a remarkably similar peak width to our own. That the peaks of V2021 are markedly wider would suggest that a significant proportion of sources in their stacks appear as resolved at Phase I resolution. Moreover, the lack of a `stepped peak', caused by the addition of a dominant unresolved population and a fainter resolved population, would suggest that the resolved population actually dominates in the V2021 stacks. This is a fundamental discrepancy with our own results, for which we do not currently have an explanation.
 
\subsection{The effect of CLEANing}
\label{sec:cleaning}

\begin{figure}
    \centering
    \includegraphics[width=\linewidth,clip,trim={0.9cm 0cm 1cm 1cm}]{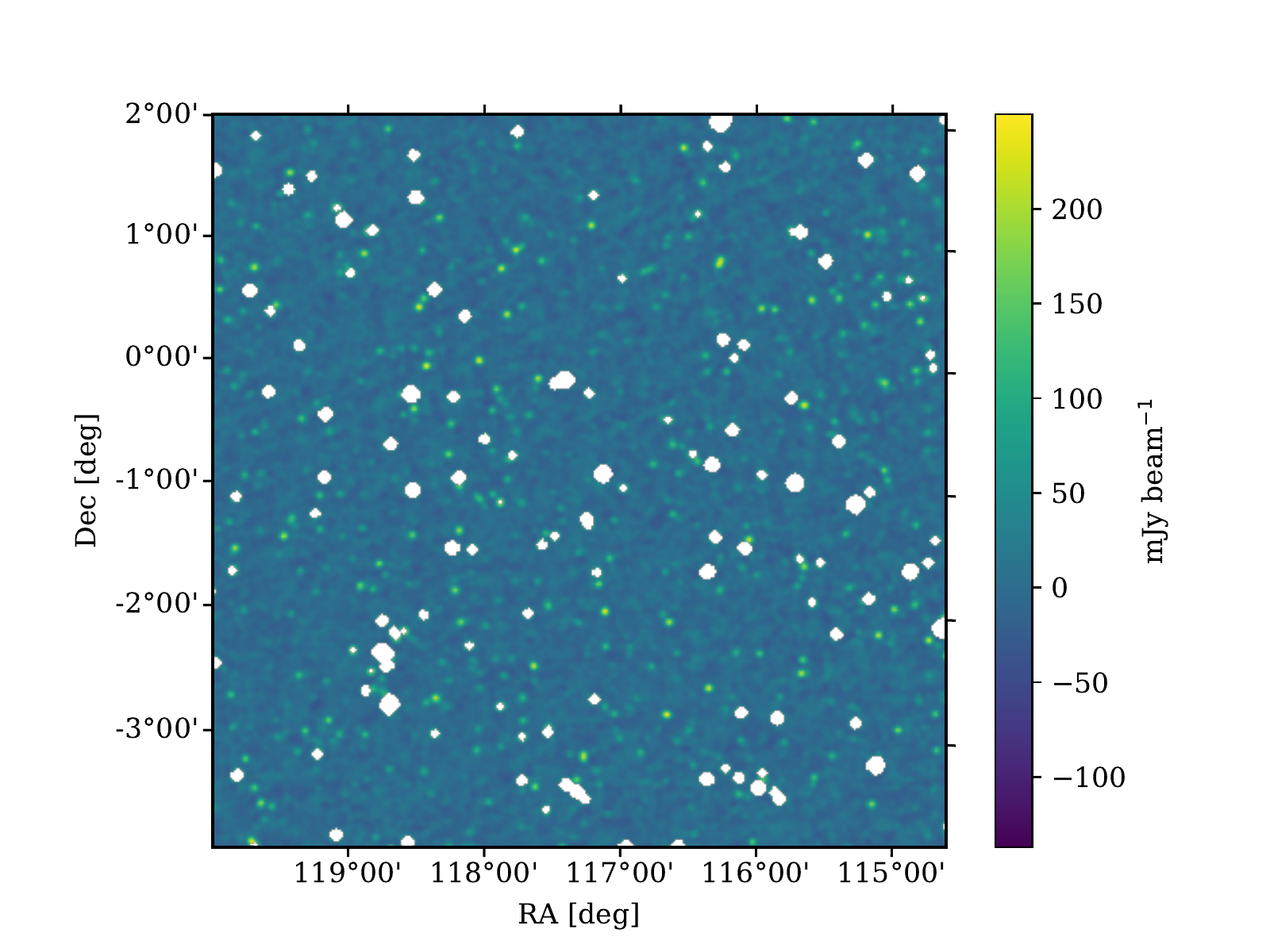}
    \caption{An example of masking the restored fields using a threshold of \SI{250}{\milli \jansky \per \beam}, with masked sources depicted here as white. Note the presence of a low to medium brightness population of radio sources still clearly visible.}
    \label{fig:maskedfield}
\end{figure}

\begin{figure*}
    \centering
    \begin{subfigure}{\linewidth}
        \centering
        \includegraphics[width=0.7\linewidth,clip,trim={2cm 1cm 0cm 2cm}]{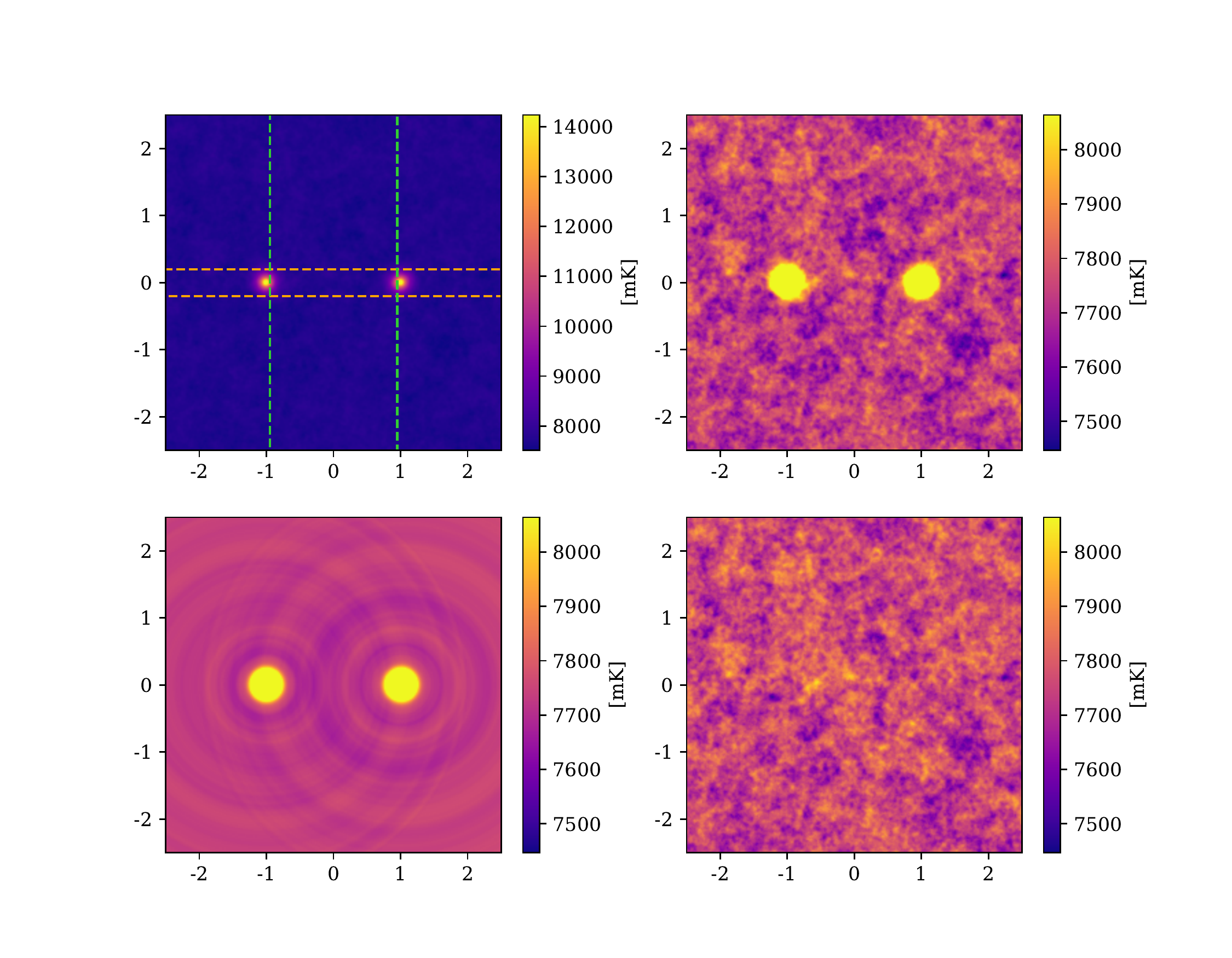}
        \caption{\textit{Top left:} The original mean stack image, with overlays indicating the region over which the transverse mean (dashed orange horizontal lines) and longitudinal mean (dashed green vertical lines) are calculated. \textit{Top right:} The mean stacked image with the colour scale reduced to $\pm 5 \sigma$ to emphasise the noise. \textit{Bottom left:} The model image, on the same colour scale. \textit{Bottom right:} The residual stack after model subtraction, with the colour scale set to $\pm 5 \sigma$.} 
    \end{subfigure}
    \begin{subfigure}{\linewidth}
        \centering
        \includegraphics[width=0.8\linewidth,clip,trim={0 0 0 -0.5cm}]{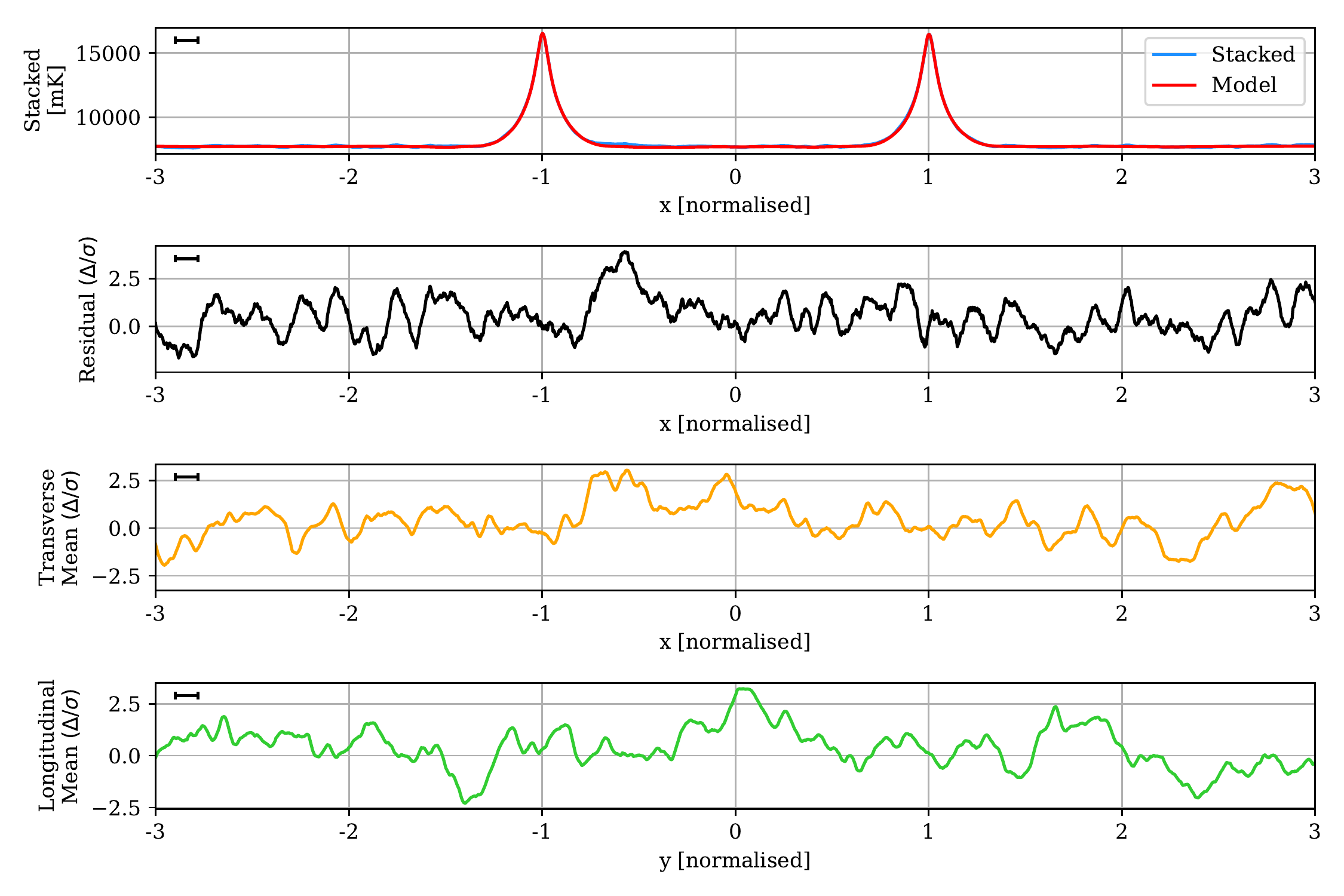}
        \caption{\textit{One:} The one-dimensional profile along $y = 0$ for both the stacked image (blue) and the model (red). \textit{Two:} The one-dimensional profile along $y = 0$ of the residual stack, renormalised to the estimated map noise. \textit{Three:} The transverse mean along the region $-0.2 < y < 0.2$ of the residual stack, renormalised to the estimated map noise. \textit{Four:} The longitudinal mean along the region $-0.95 < x < 0.95$ of the residual stack, renormalised to the estimated map noise. The black rule in the top left shows the FHWM of the effective resolution.}
    \end{subfigure}
    \caption{The Max \SI{15}{\mega \parsec} stack after masking fields at a threshold of \SI{250}{\milli \jansky \per \beam}, with mean LRG peaks of \SI{8776}{\milli \kelvin} above the background, residual noise of \SI{64}{\milli \kelvin}, and effective resolution of 0.12.}
    \label{fig:masked}
\end{figure*}

One point of difference between V2021 and the present study is the technique used to subtract bright point sources. V2021 used a wavelet decomposition technique, whereby image features on small angular scales were identified by imaging a limited range of wavelet scales. These small-scale image maps were searched for all pixels having values greater than $5 \sigma$ of the map noise, which were then subtracted from the original maps. This technique subtracted out the brightest pixels of point sources but left a residual ring around the sources at values lower than $5 \sigma$. Only compact, point-like sources were subtracted from the images, thus leaving extended sources; it's unclear what kind of additional filtering was applied to extended sources such as AGN lobes as this is not documented.

This differs with the present technique of using residuals after cleaning. Our cleaning process uses \texttt{wsclean} and its auto-mask and auto-threshold functionality. This worked by cleaning peaks of emission that are brighter than $3 \sigma$, and when this was exhausted, cleaning was allowed to continue within a mask defined by the existing clean components down to a level of $1 \sigma$. Recall that multiscale cleaning was disabled, and so this process removed \textit{peak} emission that was greater than $3 \sigma$; large, diffuse extended emission that did not peak above this threshold remained in the image. In practice, typical snapshot noise was approximately \SI{20}{\milli \jansky \per \beam}, meaning that peak emission fainter than approximately \SI{60}{\milli \jansky \per \beam} was left in the images. Compare this to the $5 \sigma$ threshold used in V2021, which corresponds to approximately \SI{175}{\milli \jansky \per \beam}. Thus there is significantly more emission remaining in the images of V2021.

For these point source subtraction differences to contribute to the detection in V2021, this would imply that the excess emission arises from a population of especially bright sources that are visible in our own mosaics at levels of greater than \SI{60}{\milli \jansky \per \beam} and which have been partially cleaned. \citet{Hodgson2020} showed that the luminosity of accretion shocks around the periphery of dark matter halos throughout their simulated cosmic web approximated a power law as a function of dark matter halo mass; in their \SI{15 x 15}{\degree} simulated field, there existed a few bright points of cosmic web emission, with the brightest at \SI{64}{\milli \jansky \per \beam} (using a Phase I MWA beam). Note, however, that these sources were located around the periphery of bright clusters, not in the true intercluster region, that they were morphologically akin to radio relics, and were likely stationary accretion shocks around massive clusters. Only a handful of such bright, outlier emission sources were predicted as part of the simulation.

To investigate this further, as an exercise we have re-run the stacking process using the restored field images, rather than the residuals. To mitigate the effects of bright point source emission, we have masked bright sources, but note that we have extended the threshold for this masking out to \SI{250}{\milli \jansky \per \beam}. The motivation for this much higher threshold is to capture emission that is present in the original V2021 images, but which we have removed by our deeper cleaning. \autoref{fig:maskedfield} shows an example of one of these masked fields, where we can clearly see a large population of sub-\SI{250}{\milli \jansky} sources still present.

We show the stacked results of this exercise in \autoref{fig:masked}. Firstly, note that the mean residual value is much greater than zero. This results from the significant number emission sources present in the image when masking to only a threshold of \SI{250}{\milli \jansky \per \beam}, and this non-zero background represents a kind of mean, stacked background temperature. Despite this, the peaks at $x = \{-1, 1\}$ have almost doubled against this background temperature, when compared with the Max 15 Mpc stack in \autoref{fig:max15}, showing that there is a considerable number of LRG sources (or sources otherwise correlated with the LRG population) with a peak brightness greater than approximately \SI{60}{\milli \jansky \per \beam}. As a side-effect of the number of sources remaining in the image, however, the noise has also increased compared to the Max 15 Mpc stack, by a factor of just over 2.5 times at \SI{62}{\milli \kelvin}. Note also the absence of the negative sidelobes about the LRG peaks. The extra emission of the LRG peaks compared to the original Max 15 Mpc stacks is the result of restored emission that has been convolved with an elliptical Gaussian fitted to the dirty beam, and this additional component will not have sidelobes; these brighter Gaussian sources in the stacks, combined with the overall higher noise, have washed out the subtle sidelobes of the fainter, uncleaned sources.

Turning now to the detection of excess emission, we can observe in \autoref{fig:masked} that there is a peak of emission in the residual image centred at $(x, y) = (-0.57, 0.035)$, slightly off the $y$-axis, and peaking at $4.58 \sigma$ significance. This peak corresponds to the peak in the one-dimensional profile also at $x = -0.57$. The width of the peak is slightly extended beyond the FWHM typical of the rest of the residual image. A second, smaller peak is also evident in the residuals at $(x, y) = (-0.07, 0.13)$ with $4.2 \sigma$ significance, and is also visible in the transverse mean. Combined, these two peaks contribute to a peak in the longitudinal mean, at $y = 0.04$ with $3.2\sigma$ significance. Note also the presence of a $4.1 \sigma$ peak outside and to the left of the stacked intercluster region, at $(x, y) = (-1.85, 0.17)$.

Are these emission peaks in the stacked residuals of \autoref{fig:masked} evidence of the cosmic web? We can immediately note that these emission peaks have not reproduced the broad, excess emission of the kind in V2021 that filled the intercluster bridge; instead these are much more localised peaks. We can also note the asymmetry of the left peak at $x = -0.57$, which is not reproduced on the right: this would suggest that this is not a generalised feature of the intercluster region. Additionally, the $4.1 \sigma$ peak to the left of the intercluster region cannot, by its location, be attributable to intercluster cosmic web emission. To investigate further, we have jackknife sampled the Max 15 Mpc catalogue, excluding a randomly selected 10\% of the catalogue, and stacked each of the ten sub-catalogues. With 90\% of the original catalogue in each stack, the noise is very similar, varying between \SIrange[range-phrase=\ to\ ,range-units=single]{63}{65}{\milli \kelvin}. We find that the peak at $x = -0.6 \pm 0.1$ is present in each stack, with at least a significance of $3.1 \sigma$, with the exception of one of the sub-catalogues, where it is entirely consistent with the noise, and peaking at most at $2.4 \sigma$. Similarly, the peaks at $x = -0.07$ and $x = -1.85$ are also each absent in one of the sub-catalogues. This exercise suggests that these peaks are not generalised features shared across the sample, but the effect of bright outlier emission left in the original fields.

This exercise suggests the absence of the broad excess emission feature found in V2021 in our own stacks is not a side-effect of cleaning. 

% Another consideration is the effect of the exclusion zones we constructed prior to stacking, and specifically the ad hoc exclusion zones constructed by visually identifying imaging artefacts or extended sources. It is not clear whether V2021 excluded such sources from stacking as it is not documented in their paper. Thus the possibility exists that among our exclusion zones are bright, outlier emission from the cosmic web that is giving rise to the signal detected in V2021. However, we can test for this simply by stacking our fields without first erasing the exclusion zones. And indeed, for each LRG catalogue we find stacked signals that are visually similar to those of their exclusion counterparts, suggesting that emission in these zones has minimal effect on the overall stacked signals.

\subsection{Stacking the original GLEAM survey}

\begin{figure*}
    \centering
    \begin{subfigure}{\linewidth}
        \centering
        \includegraphics[width=0.7\linewidth,clip,trim={2cm 1cm 0cm 2cm}]{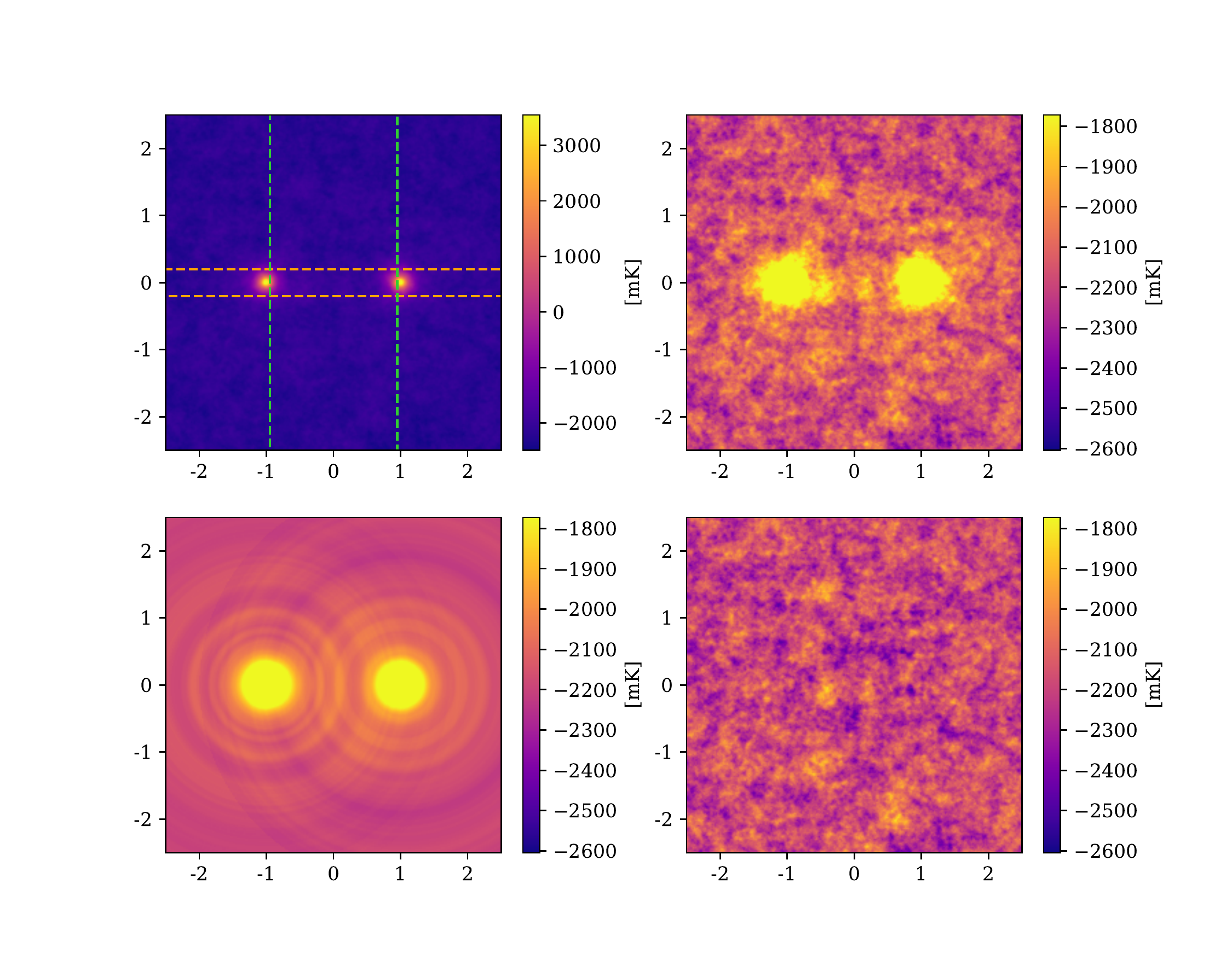}
        \caption{\textit{Top left:} The original mean stack image, with overlays indicating the region over which the transverse mean (dashed orange horizontal lines) and longitudinal mean (dashed green vertical lines) are calculated. \textit{Top right:} The mean stacked image with the colour scale reduced to $\pm 5 \sigma$ to emphasise the noise. \textit{Bottom left:} The model image, on the same colour scale. \textit{Bottom right:} The residual stack after model subtraction, with the colour scale set to $\pm 5 \sigma$.} 
    \end{subfigure}
    \begin{subfigure}{\linewidth}
        \centering
        \includegraphics[width=0.8\linewidth,clip,trim={0 0 0 -0.5cm}]{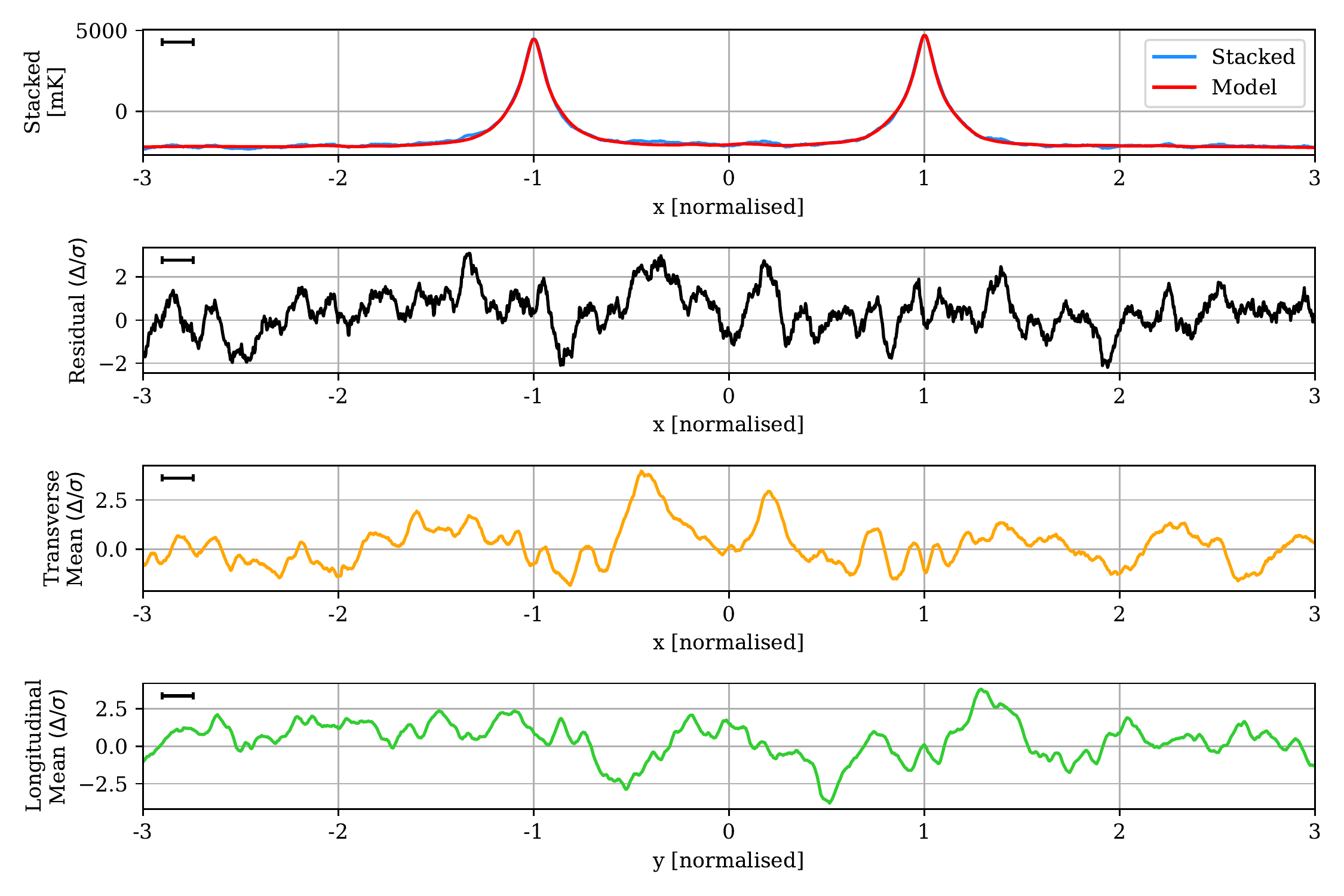}
        \caption{\textit{One:} The one-dimensional profile along $y = 0$ for both the stacked image (blue) and the model (red). \textit{Two:} The one-dimensional profile along $y = 0$ of the residual stack, renormalised to the estimated map noise. \textit{Three:} The transverse mean along the region $-0.2 < y < 0.2$ of the residual stack, renormalised to the estimated map noise. \textit{Four:} The longitudinal mean along the region $-0.95 < x < 0.95$ of the residual stack, renormalised to the estimated map noise. The black rule in the top left shows the FHWM of the effective resolution.}
    \end{subfigure}
    \caption{Original GLEAM survey images at \SI{118.5}{\mega \hertz}, stacked using the Max \SI{15}{\mega \parsec} LRG catalogue, displaying mean LRG peaks of \SI{4600}{\milli \kelvin}, residual noise of \SI{87}{\milli \kelvin}, and effective resolution of 0.16.}
    \label{fig:GLEAMstacks}
\end{figure*}

We have every expectation that we should be able to detect the excess emission in our Phase II observations, given the low noise characteristics of our fields, our sensitivity to large-scale angular structures, and the additional LRG pairs that we have used in our stacks. It is still possible, however, that there is some aspect of these new observations or our image processing pipelines that has obscured or removed the synchrotron cosmic web. And so these concerns have led us to return to the original GLEAM survey data, and attempt to reproduce the results of V2021 by stacking an identical data set at \SI{118.5}{\mega \hertz}.

To stack the GLEAM survey data, we first start with the full zenith equal area (ZEA) projection images at \SI{118.5}{\mega \hertz}, which cover the right ascension regions spanned by our LRG pairs. Unlike V2021, we leave these images in their original projection. We mask bright points by selecting all emission regions with values greater than $5 \sigma$ of the local noise. To do this, we measure the spatially variable background---which is primarily the result of Galactic emission---as well as the noise using the Background and Noise Estimation tool \citep[BANE;][]{Hancock2018}. The mask is then created by subtracting out the background emission from the full projection, dividing by the noise image, and then masking all regions that exceed $5 \sigma$ of the local noise value. This process is substantially simpler than the wavelet subtraction method used in V2021, and the inclusion of the background subtraction step mitigates their stated concerns about masking. We additionally `grow' all masked regions by 2 pixels, which we have found to be sufficient to avoid visually obvious rings of faint emission around the masks. Note that even after growing the masks slightly, this process leaves surrounding negative sidelobes about the masked regions, and this results in the remaining non-masked region having an overall negative mean. As with the previous masked stacks in \autoref{sec:cleaning}, this will affect the `zero point' of the final stacks. As previously, exclusion zones are identified around exceptionally bright sources, along sidelobe artefacts, and in one additional region where the background estimation had not been adequate due to a sharp change in the background brightness. Finally, the map is converted to temperature using the associated dirty beam map that described the major and minor axis variation of the beam. We then proceed to stack all LRG pairs from the Max \SI{15}{\mega \parsec} catalogue that overlap with the images, of which there are 645,950 unique pairs. All stacks are weighted by the inverse square of the local noise.

In \autoref{fig:GLEAMstacks}, we show the results of the GLEAM stacking with the Max \SI{15}{\mega \parsec} catalogue. This residual image is noticeably different from the one presented in V2021. In the first instance, there is no obvious, large-scale region of excess emission. In V2021, this excess region spanned the length of the intercluster region, and surprisingly was wider than it was long. The residual image is also highly uniform, again differing from V2021 where all the residual images, including the null tests, displayed a distinctive large scale pattern. Curiously, the noise in this image is at \SI{87}{\milli \kelvin}, which is higher than reported in V2021 even though we stack a much larger number of LRG pairs.\footnote{Note that the stated value in V2021 was calculated assuming the average noise in the original images reduced from stacking $N$ LRG pairs as a factor of $1 / \sqrt{N}$, rather than being measured directly.} The one-dimensional profiles similarly display little evidence of excess emission in the residual, with the exception of a peak that reaches $3 \sigma$ in the integrated profile, at $x \approx -0.5$, and which has a width very slightly wider than the effective resolution. There are at least 3 other peaks of similar magnitude and size throughout the residual image that cannot be attributed to intercluster cosmic web emission by reason of their location in the map, and so we must conclude that this peak is unexceptional.

Ultimately, we are unable to reproduce the broad and extended excess emission signal found in V2021, even when using the same data set, raising questions that these differences in results are due to the stacking procedure. In \autoref{sec:rosat}, we perform a similar stacking procedure on the ROSAT broad X-ray data, as was performed in V2021. In this case, however, we detect a strong $12 \sigma$ signal for the Max \SI{15}{\mega \parsec} catalogue. This confirms the detection of V2021 for this data set, and provides us with confidence that our stacking and model subtraction processes will detect excess emission when it is present, and suggests that the discrepancy in results arises elsewhere in the analysis.

\section{Conclusion}

We have attempted to reproduce the detection of excess emission spanning LRG pairs in low frequency radio data, as reported by V2021, and which they attributed to synchrotron emission along filaments spanning pairs of close-proximity clusters and galaxy groups. To reproduce their work, we have adhered very closely to their methodology: using the same LRG catalogue and selection criteria for pairs, stacking radio images at \SI{118.5}{\mega \hertz}, and modelling the LRG and cluster contribution in the same way as V2021. We differ from V2021 primarily in that we use the upgraded MWA Phase II array, which has almost twice the resolution as the Phase I instrument used in V2021, and that our calibration, imaging and point-source subtraction pipelines utilised improved workflows that have been developed since the original GLEAM survey.

We have not been able to reproduce their result. Indeed, we have not been able to reproduce their result across a number of LRG pair catalogues, including the original abridged catalogue used in V2021, as well as a much larger catalogue that uses the full range of LRG pairs that meet the original selection criteria of V2021. We reach noise levels in our final stacks consistently lower than those of V2021, and more than twice as deep when using the full range of available LRG pairs. At these noise levels, their reported filamentary temperature should appear as approximately an $8 \sigma$ detection. Our residual stacks, however, are consistent with noise.

Our biggest concern with using MWA Phase II is the potential that we resolve out large, extended structures. However, we have shown that we are at least as sensitive to extended sources as Phase I out to $\sim$\SI{125}{\arcminute} thanks to our extra convolution step, and that even for extended emission up to the maximum separation of \SI{180}{\arcminute}, the likely shape and structure of this emission will reduce the effects of resolving out structure. Moreover, we have provided results of an additional LRG pair catalogue, with sources separated by \SIrange[range-phrase=\ to\ ]{15}{60}{\arcminute}, that mitigates these concerns; the stacking results of this catalogue reach noises lower than those of V2021 and yet still do not reproduce their observed excess emission.

In addition, we have returned to the original GLEAM survey data where we have performed stacking using the expanded Max \SI{15}{\mega \parsec} LRG pair catalogue. Whilst we do find an isolated peak at just above $3 \sigma$ significance, we find this to be an unremarkable feature of the residuals and certainly not the broad, extended excess emission as found in V2021. This non-detection is in spite of clearly reproducing the excess emission after stacking the ROSAT broad X-ray data, giving us good confidence in our stacking and modelling processes.

If our results hold true, we have provided in this work the strongest limits on synchrotron emission from intercluster filaments. However, the discrepancy with the work of V2021 is concerning and begs explanation. Whilst our Phase II results alone left open the possibility that this discrepancy arose due to a real, intrinsic property of the emission, our inability to reproduce the results additionally with GLEAM points to a much more likely possibility: that an error has been made in these detection attempts either by V2021 or ourselves. To this end, we are making publicly available the images of our fields, our stacking and modelling code, and the stacked images themselves, in the hope that if we have indeed erred, it can be quickly identified. Given the significance of the V2021 result, and the surprising implications on our understanding of cosmic magnetism, there is a pressing need reproduce their detection.

\section*{Acknowledgements}

This scientific work makes use of the Murchison Radio-astronomy Observatory, operated by CSIRO. We acknowledge the Wajarri Yamatji people as the traditional owners of the Observatory site. Support for the operation of the MWA is provided by the Australian Government (NCRIS), under a contract to Curtin University administered by Astronomy Australia Limited. We acknowledge the Pawsey Supercomputing Centre which is supported by the Western Australian and Australian Governments.

This work was supported by resources provided by the Pawsey Supercomputing Centre with funding from the Australian Government and the Government of Western Australia.

We acknowledge the use of NASA's SkyView facility (\url{http://skyview.gsfc.nasa.gov}) located at NASA Goddard Space Flight Center.

%%%%%%%%%%%%%%%%%%%% REFERENCES %%%%%%%%%%%%%%%%%%

% The best way to enter references is to use BibTeX:

\bibliographystyle{pasa-mnras}
\bibliography{main} % if your bibtex file is called example.bib

% Alternatively you could enter them by hand, like this:
% This method is tedious and prone to error if you have lots of references
%\begin{thebibliography}{99}
%\bibitem[\protect\citeauthoryear{Author}{2012}]{Author2012}
%Author A.~N., 2013, Journal of Improbable Astronomy, 1, 1
%\bibitem[\protect\citeauthoryear{Others}{2013}]{Others2013}
%Others S., 2012, Journal of Interesting Stuff, 17, 198
%\end{thebibliography}

%%%%%%%%%%%%%%%%%%%%%%%%%%%%%%%%%%%%%%%%%%%%%%%%%%

%%%%%%%%%%%%%%%%% APPENDICES %%%%%%%%%%%%%%%%%%%%%

\appendix

\section{Synthetic stacks}
\label{sec:synthetic}

\begin{figure}
    \centering
    \includegraphics[width=0.8\linewidth,clip,trim={1.5cm 0.8cm 1.5cm 0.5cm}]{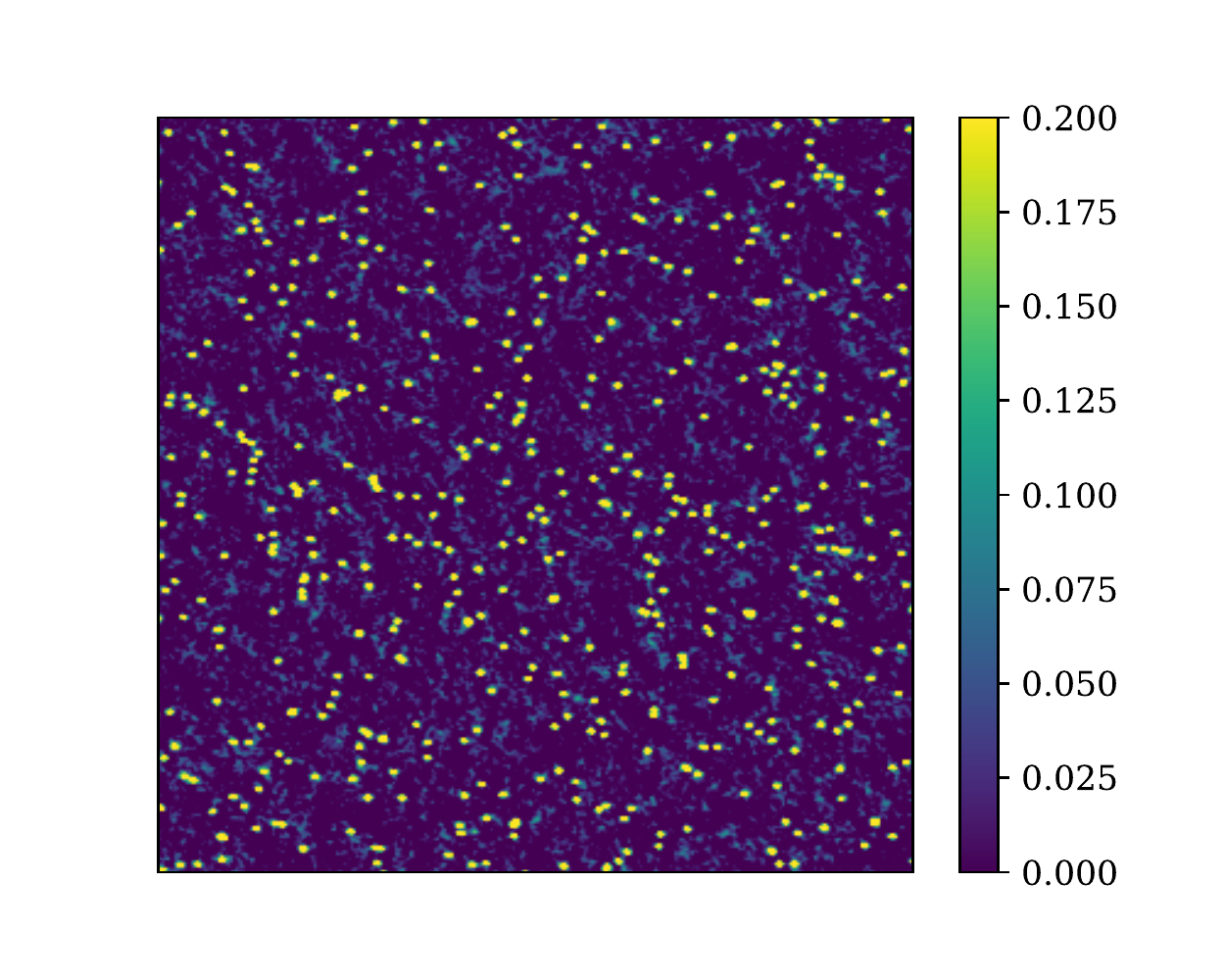}
    \caption{A zoom of the synthetic cosmic web image, showing randomly positioned \SI{1}{\kelvin} point sources with a subset of pairs connected by faint `filaments'. After convolution to with the Phase II (convolved) beam, the filaments become hidden beneath the \SI{25}{\milli \kelvin} sidelobe confusion noise.}
    \label{fig:syntheticmaps}
\end{figure}

\begin{figure*}
    \centering
    \begin{subfigure}{\linewidth}
        \centering
        \includegraphics[width=0.7\linewidth,clip,trim={2cm 1cm 0cm 2cm}]{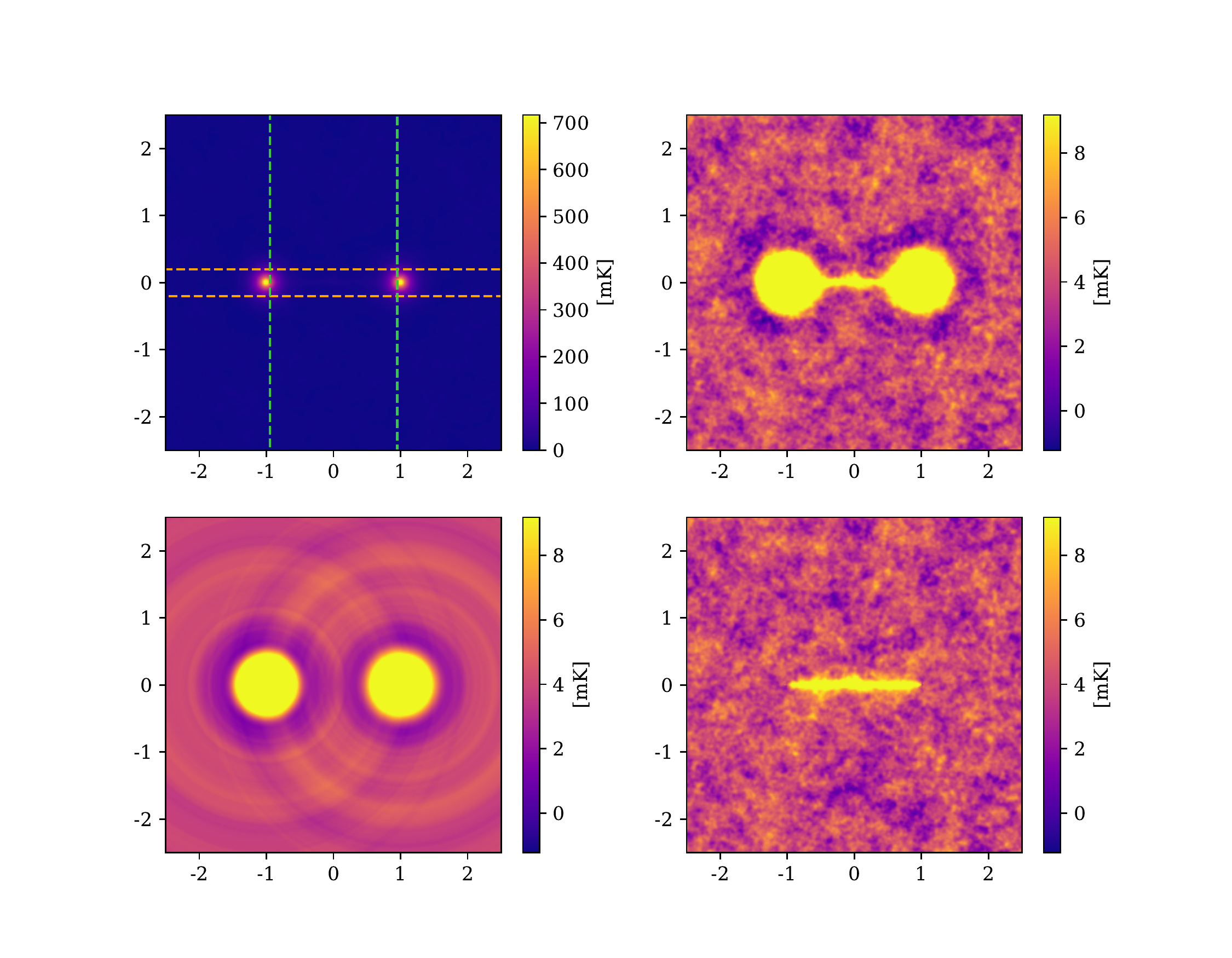}
        \caption{\textit{Top left:} The original mean stack image, with overlays indicating the region over which the transverse mean (dashed orange horizontal lines) and longitudinal mean (dashed green vertical lines) are calculated. \textit{Top right:} The mean stacked image with the colour scale reduced to $\pm 5 \sigma$ to emphasise the noise. \textit{Bottom left:} The model image, on the same colour scale. \textit{Bottom right:} The residual stack after model subtraction, with the colour scale set to $\pm 5 \sigma$.} 
    \end{subfigure}
    \begin{subfigure}{\linewidth}
        \centering
        \includegraphics[width=0.8\linewidth,clip,trim={0 0 0 -0.5cm}]{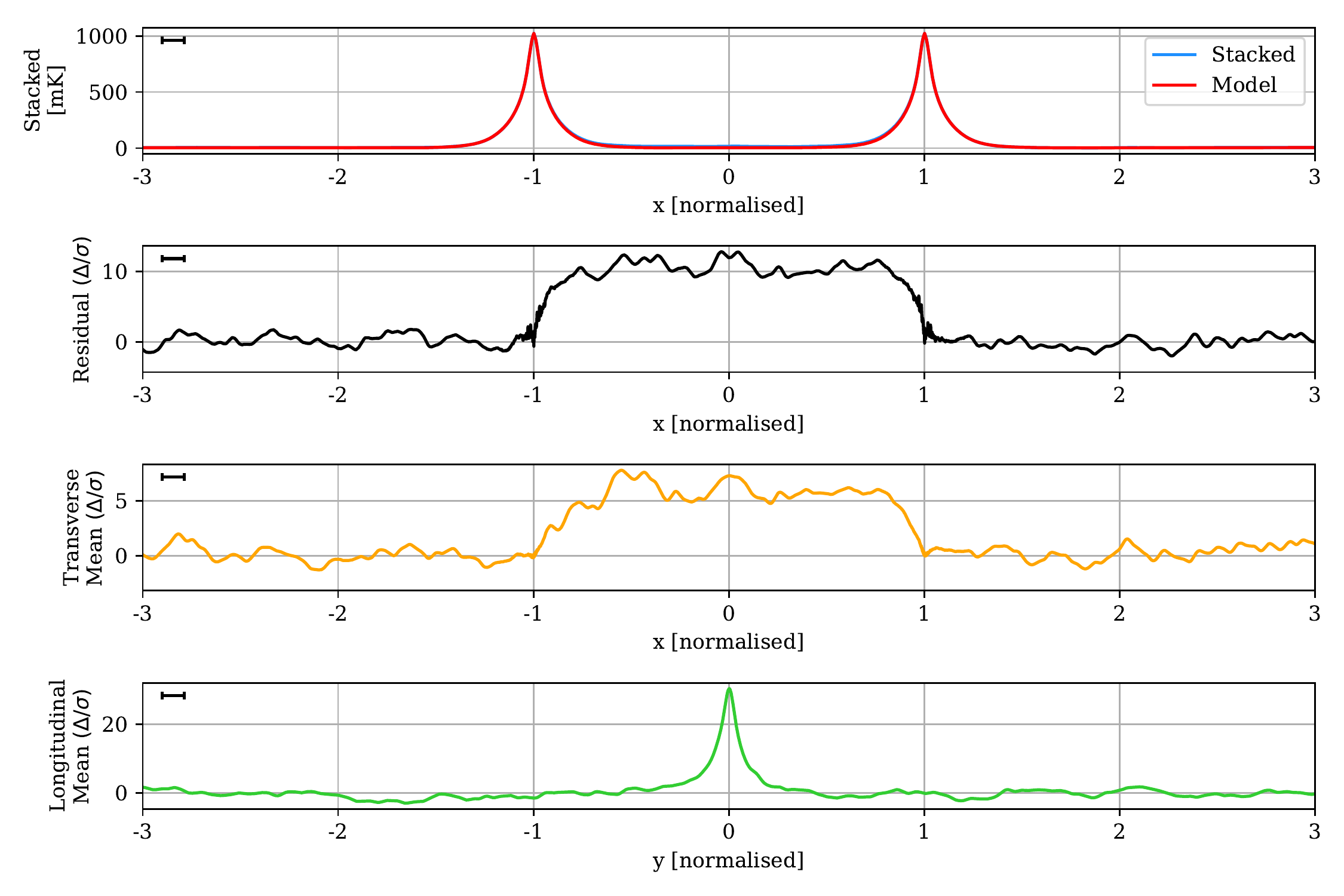}
        \caption{\textit{One:} The one-dimensional profile along $y = 0$ for both the stacked image (blue) and the model (red). \textit{Two:} The one-dimensional profile along $y = 0$ of the residual stack, renormalised to the estimated map noise. \textit{Three:} The transverse mean along the region $-0.2 < y < 0.2$ of the residual stack, renormalised to the estimated map noise. \textit{Four:} The longitudinal mean along the region $-0.95 < x < 0.95$ of the residual stack, renormalised to the estimated map noise. The black rule in the top left shows the FHWM of the effective resolution.}
    \end{subfigure}
    \caption{The synthetic stacks, with mean peaks of \SI{1.02}{\kelvin}, residual noise of \SI{1}{\milli \kelvin}, and effective resolution of 0.11.}
    \label{fig:syntheticstacks}
\end{figure*}

To validate our stacking and modelling process, we have created a synthetic `cosmic web' map and associated catalogue. We created a $14000 \times 14000$ pixel map, randomly positioned 26,000 point sources with peak temperature of \SI{1}{\kelvin} and then selected pairs of peaks that were separated in the range \SIrange[range-phrase=--,range-units=single]{20}{180}{\arcminute}. Pairs were randomly selected to produce a pair separation distribution that strongly favoured shorter separations, so as to approximate the distributions as seen in \autoref{fig:lrgpairs}; 5,539 pairs were chosen in this way. For each chosen pair, we drew a `filament' as a straight, single pixel line, with each pixel having a value of \SI{0.005}{\kelvin}. Finally, we convolved the image with the Phase II (convolved) beam, as shown in \autoref{fig:psfs}. The resulting map is shown in \autoref{fig:syntheticmaps} where the point sources can be readily observed whilst the cosmic web filaments are not visually detectable above the sidelobe confusion, which is approximately \SI{25}{\milli \kelvin}.

In \autoref{fig:syntheticstacks} we show the results after stacking and model subtraction. The stacked point sources peak at approximately \SI{1}{\kelvin}, in agreement with their injected values. Note also the presence of the faint, negative lobes about the exterior sweep of each peak, resulting from the negative lobes of the dirty beam. In practice, both the width of the peaks and positioning of the sidelobes are highly dependent on the distribution of the angular separation of pairs, which in turn affects the mix of rescaling that is required during stacking; for catalogues with more distant pairs, the peaks became narrower and the sidelobes less prominent. As can be seen in the residual image, after model subtraction we clearly recover the `cosmic web' bridge. The model subtraction process does well to compensate for the negative sidelobes, resulting in a bridge that is fairly constant as can be seen in the one-dimensional profile, at approximately $11 \sigma$ in the residual along $y = 0$. This is reduced in the transverse mean profile as a result of the narrowness of the filament, however the signal exceeds $30 \sigma$ in the longitudinal profile.

We believe these results validate our stacking and modelling processes.

\section{The weighting of the stacks}
\label{sec:weighting}

\begin{figure}
    \centering
    \includegraphics[width=\linewidth,clip,trim={1.3cm 0 1.7cm 1.3cm}]{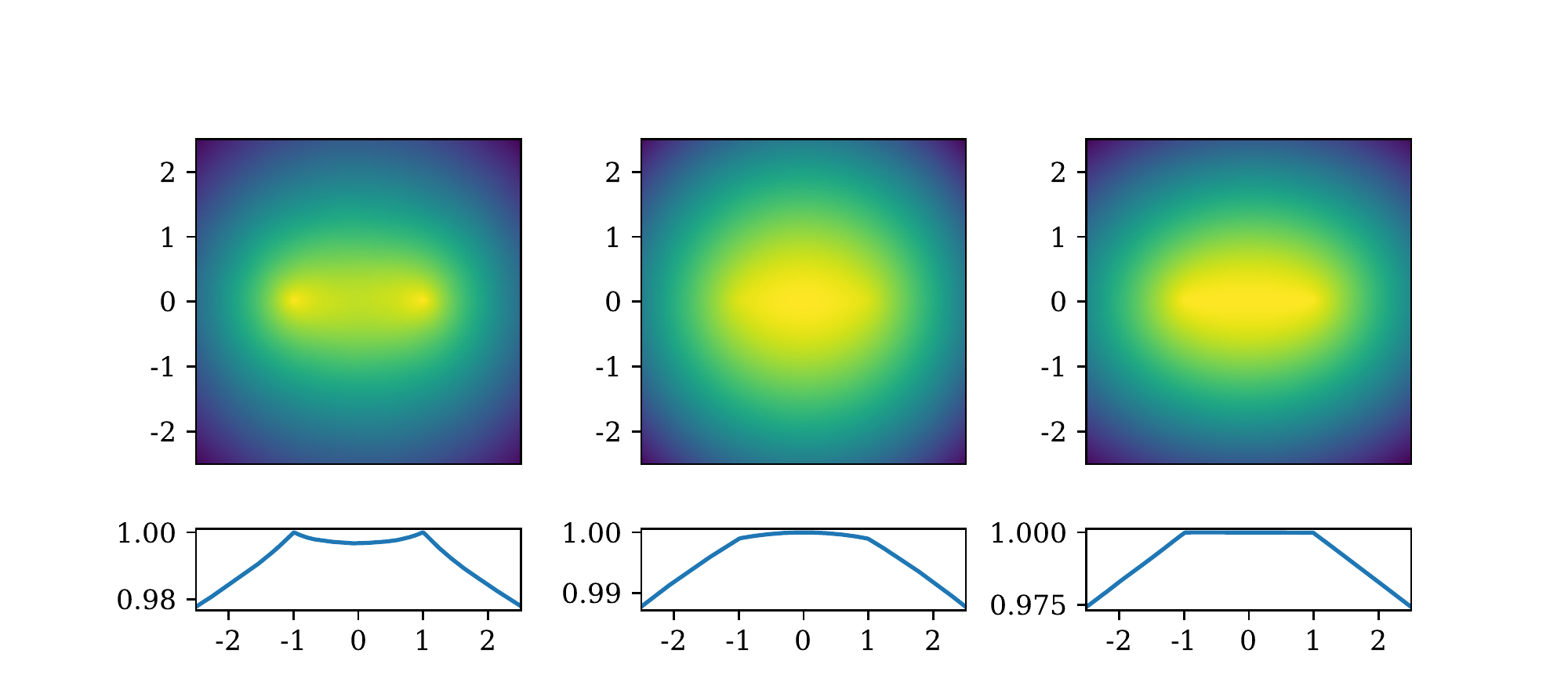}
    \caption{Stacked weight maps for the Max 15 Mpc catalogue, using a variety of weighting configurations. The weights are ultimately derived from the estimated noise maps of each field, which are spatially varying across the fields. The dominant effect seen here is the result of LRG pairs near the edge of the fields that produces the tapering effect towards the edges, with secondary effects caused by the ad-hoc exclusion zones and the convex geometry of the field perimeter. \textit{Left:} The stacked weight map resulting from the default, noise-weighted stacking, also showing a one-dimensional profile through $y = 0$ in the lower panel. \textit{Centre:} The stacked weight map using noise-weighted stacking but ad-hoc exclusion zones ignored during stacking and left unmasked. \textit{Right:} The stacked weight maps using constant weight during stacking, and with ad-hoc exclusion zones ignored.}
    \label{fig:stackednoise}
\end{figure}

When stacking the LRG pairs from each field, we have weighted each pair by an estimate of the field noise. We have previously noted that these noise maps are spatially varying across a field, as a function of the primary beam attenuation, and that even across a single LRG pair the noise may be varying. What effect does this have on the final weighting of each stack?

In \autoref{fig:stackednoise} we show the associated stacked weight maps constructed as part of the Max \SI{15}{\mega \parsec} stack for different weighting configurations. From left to right: the default $\nicefrac{1}{\sigma^2}$ noise weighted maps with ad-hoc exclusion zones as detailed in \autoref{sec:exclusionzones}, the same but with the ad-hoc exclusion zones ignored during stacking and left unmasked, and finally with a constant weighting and no ad-hoc exclusion zones. We also provide a one-dimensional profile for each map through $y = 0$, where we observe that each weighted map differs primarily by the weighting directly between the LRG pair.

Each stacked weight map shares a bright central component that tapers off towards the edge. This is not an effect of local variation of noise in each field, but rather is a result of the convex perimeter of each field. During stacking, any LRG pairs near the edge of this field window have significant areas of their rescaled and rotated images `outside' the field, and are therefore both set to zero and weighted as zero. The combined effect of this has resulted in the dominant tapering effect as observed. The secondary effect that we can observe in the stacked weight maps is the intercluster weighting, which varies from left to right as a shallow depression, shallow rise and a constant weight. There are two separate effects at work in creating these intercluster weightings. The first effect is the presence of the ad-hoc exclusion zones. Recall that the only requirement during stacking was that each LRG pair occupied a non-masked pixel, that is, that each LRG was interior to a field, and was not masked by an exclusion zone. This requirement did not, however, exclude cases where an ad-hoc exclusion zone existed between an LRG pair. This is the effect that dominates in the default example, causing the shallow depression in weighting between the LRG pair. If we ignore the ad-hoc exclusion zones during stacking, however, we instead obtain a stacked weight map with a slight rise in the intercluster region. This effect is caused by variations in local map noise caused by one special case: when stacking LRG pairs along the convex perimeter of the field. When stacking such pairs, the direct line between the LRG pair passes from the field perimeter towards the field interior, going from the maximum noise at the field edge towards a slightly reduced noise environment towards the interior. The effect of this is to slightly upweight the intercluster region. Finally, both of these effects are removed by considering the case of a constant weighted stack with no ad-hoc exclusion zones: we observe a constant weight between the LRG pair.

In practice, the weighting scheme has very little effect on the final stack. The constant weighted stack, which ignores both the local noise estimation and ad-hoc exclusion zones, has an estimated residual noise of \SI{27}{\milli \kelvin} in comparison to the \SI{25}{\milli \kelvin} of the noise-weighted stacks shown previously. Moreover, each of the three differently weighted stacks have visually identical residual maps.

\section{ROSAT stacks}
\label{sec:rosat}

\begin{figure*}
    \centering
    \begin{subfigure}{\linewidth}
        \centering
        \includegraphics[width=0.7\linewidth,clip,trim={2cm 1cm 0cm 2cm}]{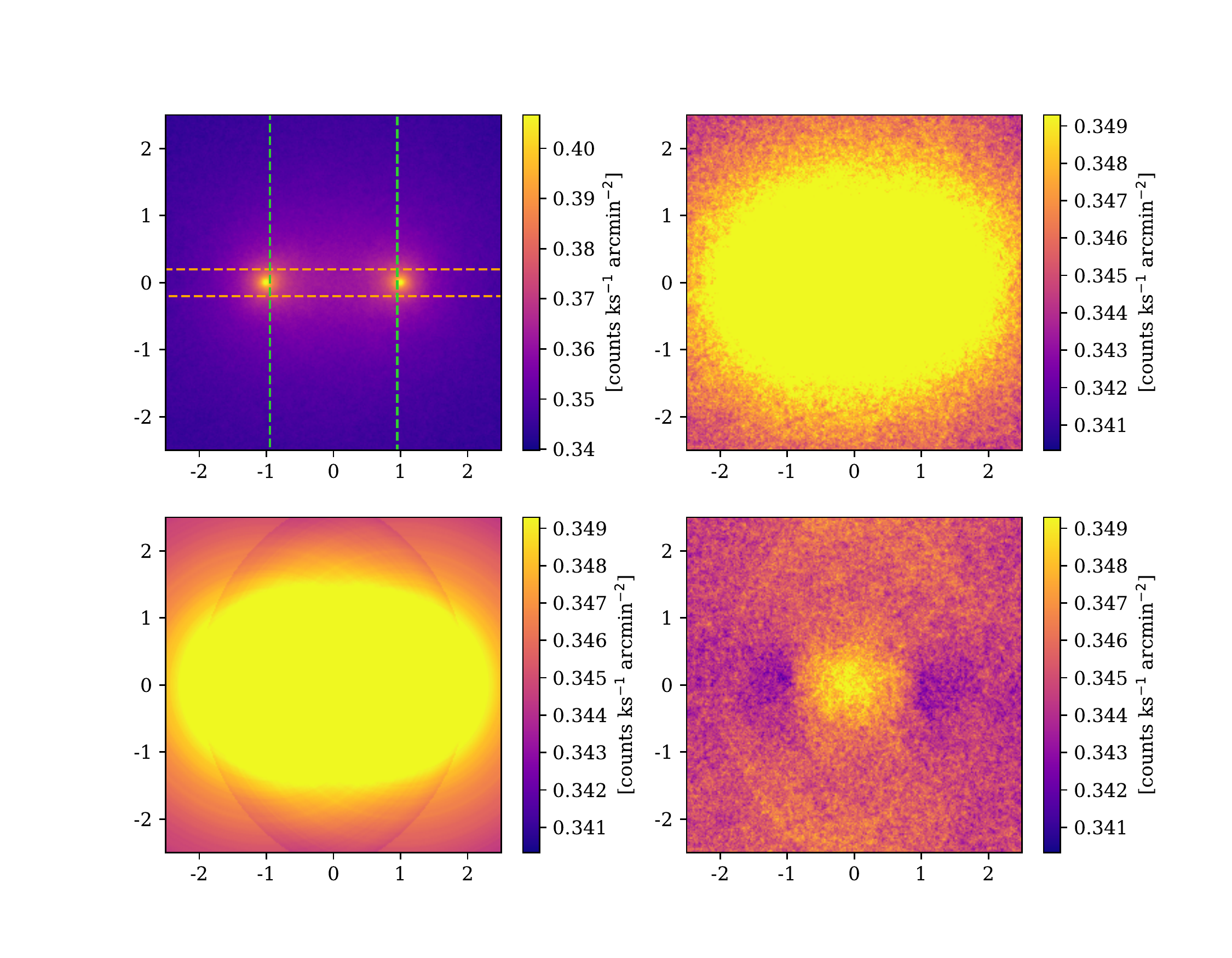}
        \caption{\textit{Top left:} The original mean stack image, with overlays indicating the region over which the transverse mean (dashed orange horizontal lines) and longitudinal mean (dashed green vertical lines) are calculated. \textit{Top right:} The mean stacked image with the colour scale reduced to $\pm 5 \sigma$ to emphasise the noise. \textit{Bottom left:} The model image, on the same colour scale. \textit{Bottom right:} The residual stack after model subtraction, with the colour scale set to $\pm 5 \sigma$.} 
    \end{subfigure}
    \begin{subfigure}{\linewidth}
        \centering
        \includegraphics[width=0.8\linewidth,clip,trim={0 0 0 -0.5cm}]{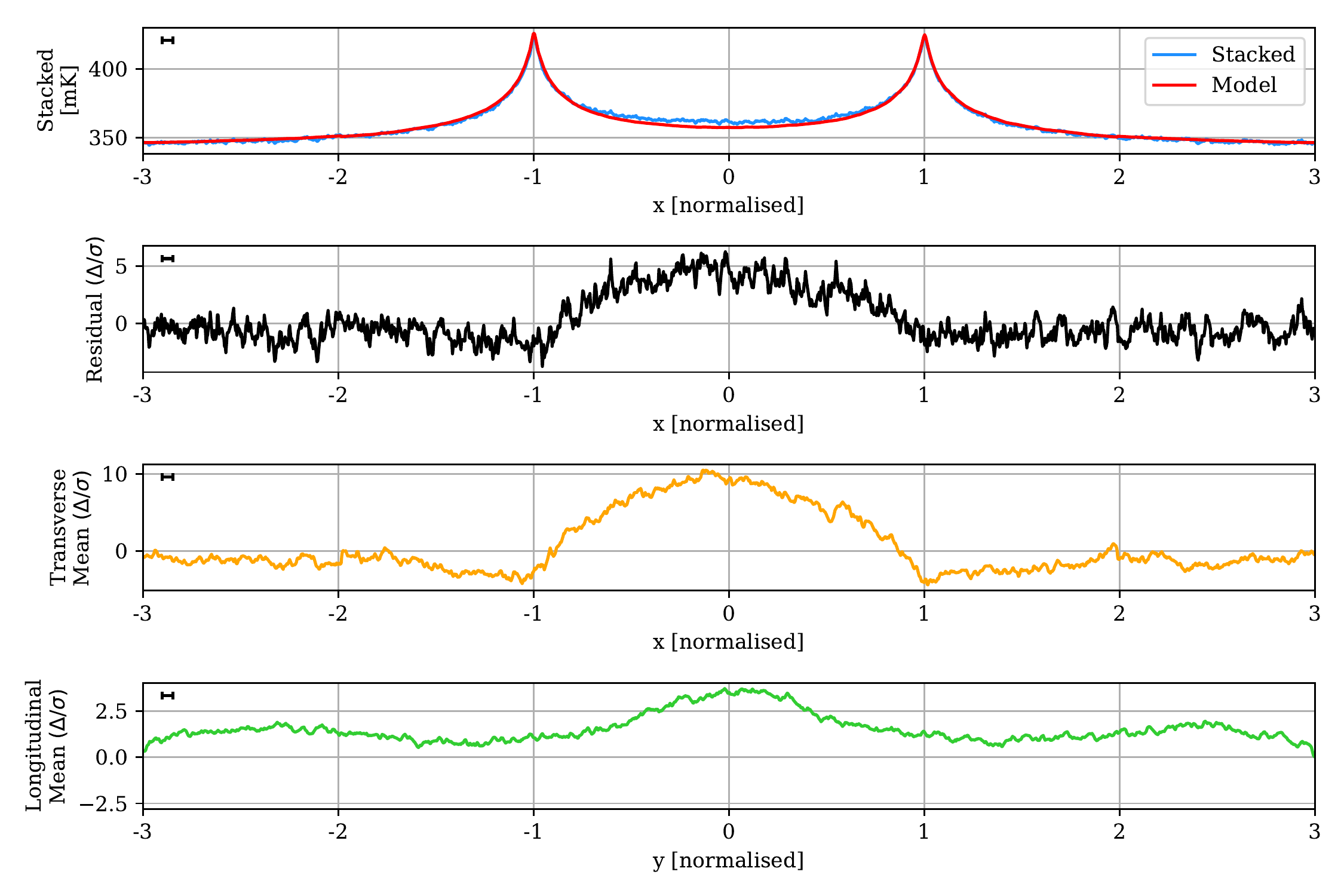}
        \caption{\textit{One:} The one-dimensional profile along $y = 0$ for both the stacked image (blue) and the model (red). \textit{Two:} The one-dimensional profile along $y = 0$ of the residual stack, renormalised to the estimated map noise. \textit{Three:} The transverse mean along the region $-0.2 < y < 0.2$ of the residual stack, renormalised to the estimated map noise. \textit{Four:} The longitudinal mean along the region $-0.95 < x < 0.95$ of the residual stack, renormalised to the estimated map noise. The black rule in the top left shows the FHWM of the effective resolution.}
    \end{subfigure}
    \caption{The Max \SI{15}{\mega \parsec} stack of ROSAT broad images, with mean LRG peaks of \SI{0.4238}{\counts \per \kilo \second \per \arcmin \squared}, residual noise of \SI{0.9E-3}{\counts \per \kilo \second \per \arcmin \squared}, and effective resolution of 0.05.}
    \label{fig:ROSATstacks}
\end{figure*}

V2021 performed stacking on X-ray data from ROSAT All-sky Survey \citep{Voges1999}, and we reproduce this here using the ROSAT broad images (\SIrange[range-phrase=--,range-units=single]{0.1}{2.4}{\kilo \electronvolt}). These images span further North than is possible with the MWA pointings, and so we are able to stack 757,731 LRG pairs as part of the Max \SI{15}{\mega \parsec} catalogue. The ROSAT broad data were downloaded as a series of \SI{20 x 20}{\degree} images in a gnomonic (TAN) projection, spanning the field of LRG pairs at intervals of \SI{10}{\degree} on the sky. Bright pixels having a count greater than \SI{20}{\counts \per \kilo \second \per \arcmin \squared} were blanked, but otherwise the images were not further processed.

The results of this stacking are shown in \autoref{fig:ROSATstacks}. The residual image shows a large, excess region centred at $x = 0$, but spanning the length of the intercluster region, and having a width approximately $-0.5 < y < 0.5$. The excess signal is very well detected, having a one-dimensional profile that peaks just above $7 \sigma$, corresponding to an excess value of \SI{6.6(9)E03}{\counts \per \kilo \second \per \arcmin \squared}, and an integrated profile (between $-0.2 < y < 0.2$) that peaks above $12 \sigma$. A null test using unrelated LRG pairs ($\Delta r >$~\SI{150}{\mega \parsec}) but otherwise conforming to the angular separation distribution of the Max \SI{15}{\mega \parsec} catalogue returned no excess signal. These excess values compare to the value found by V2021 of \SI{11.5(14)E-3}{\counts \per \kilo \second \per \arcmin \squared}, although this value was found when stacking the abridged LRG-V2021 catalogue and was a mean across an unspecified `filamentary region'.

Note that the width of X-ray emission around each LRG peak is extremely broad. In fact, in the one-dimensional profile we can see the exterior sides do not become flat even out to $x = \{-3, +3\}$. Indeed, this extreme width of each peak causes our model subtraction process to handle poorly. As we noted in \autoref{sec:modelling}, we independently model each LRG peak before simply adding each of their contributions. This works well when the contribution from each peak drops to zero for radial distances $r > 2$. However, since in this case each peak still includes a non-negligible component present from the opposing peak, we therefore over-subtract, causing both the negative bowls of emission that can be seen in the residual image, as well as to underestimate the central excess emission.

Despite detecting a signal, we would hesitate to attribute this emission to true intercluster X-ray emission. The width of each peak implies that most of the emission in the centre of the residual images originates from cluster emission, and that a significant number of these pairs must overlap along our line of sight due to projection effects. It is therefore equally plausible that the excess emission in the centre originates due to asymmetric cluster emission, rather than hot intercluster filamentary gas. This is especially plausible given the number of clusters known to host substructures away from their core \citep{Schuecker2005}.

%%%%%%%%%%%%%%%%%%%%%%%%%%%%%%%%%%%%%%%%%%%%%%%%%%

% Don't change these lines
% \bsp	% typesetting comment
% \label{lastpage}
\end{document}